\documentclass[twocolumn, twocolappendix]{aastex631}

\usepackage{amsmath}

\begin{document}

\title{A Deep Redshift Survey of the Perseus Cluster: Spatial Distribution and Kinematics of Galaxies}

\correspondingauthor{Ho Seong Hwang}
\email{hhwang@astro.snu.ac.kr}

\author{Wooseok Kang}
\affiliation{Department of Physics and Astronomy, Seoul National University, 1 Gwanak-ro, Gwanak-gu, Seoul 08826, Republic of Korea}

\author{Ho Seong Hwang}
\affiliation{Department of Physics and Astronomy, Seoul National University, 1 Gwanak-ro, Gwanak-gu, Seoul 08826, Republic of Korea}
\affiliation{SNU Astronomy Research Center, Seoul National University, 1 Gwanak-ro, Gwanak-gu, Seoul 08826, Republic of Korea}

\author{Hyunmi Song}
\affil{Department of Astronomy and Space Science, Chungnam National University, Daejeon 34134, Republic of Korea}

\author{Changbom Park}
\affil{School of Physics, Korea Institute for Advanced Study (KIAS), 85 Hoegiro, Dongdaemun-gu, Seoul, 02455, Republic of Korea}

\author[0000-0002-2013-1273]
{Narae Hwang}
\affil{Korea Astronomy and Space Science Institute, 776 Daedeokdae-ro, Yuseong-gu, Daejeon 34055, Korea}

\author{Byeong-Gon Park}
\affil{Korea Astronomy and Space Science Institute, 776 Daedeokdae-ro, Yuseong-gu, Daejeon 34055, Korea}

\begin{abstract}
    We study the global kinematics of the Perseus galaxy cluster (Abell 426) at redshift z = 0.017 using a large sample of galaxies from our new MMT/Hectospec spectroscopic observation for this cluster. The sample includes 1447 galaxies with measured redshifts within 60\arcmin\ from the cluster center (1148 from this MMT/Hectospec program and 299 from the literature). The resulting spectroscopic completeness is 67\% at r-band apparent magnitude $r_{\mathrm{Petro, 0}}\leq 18.0$ within 60\arcmin\ from the cluster center. To identify cluster member galaxies in this sample, we develop a new open-source Python package, CausticSNUpy. This code implements the algorithm of the caustic technique and yields 418 member galaxies within 60\arcmin\ of the cluster. We study the cluster using this sample of member galaxies. The cluster shows no significant signal of global rotation. A statistical test shows that the cluster does not have a noticeable substructure within 30\arcmin. We find two central regions where the X-ray emitting intracluster medium and galaxies show significant velocity differences ($>7\sigma$).
    On a large scale, however, the overall morphology and kinematics between the intracluster medium and galaxies agree well. Our results suggest that the Perseus cluster is a relaxed system and has not experienced a recent merger.
\end{abstract}


\section{Introduction}  \label{sec:intro}
    Clusters of galaxies form through the collapse of initial density fluctuations and hierarchical merging.
    Clusters are regarded as laboratories for astrophysical phenomena, such as high-energy physics and the evolution of galaxies \citep{Allen...2011ARA&A..49..409A}.
    Thus, studies of galaxy clusters and their components give insight into both cosmology and astrophysics.
    The constituents of galaxy clusters are dark matter, galaxies, and intracluster medium (ICM).
    The mass fractions of the three components are 80--95\% for dark matter, 5--20\% for ICM, and 0.5--3\% for galaxies \citep{Ettori...2009A&A...501...61E}. 
    
    Different mechanisms are applied to each component.
    Dark matter and the stellar component of galaxies are considered collisionless, even though sometimes interactions between galaxies play a key role in galaxy evolution \citep[e.g.,][]{Park...2009ApJ...699.1595P, Hwang...2012A&A...538A..15H}. The ICM gas is the collisional component of the cluster, governed not only by gravity but also by magnetohydrodynamics and thermodynamics.
    Because of this difference, dark matter and galaxies have relaxation timescales different from the gas after a cluster experiences a merger \citep{Roetigger...2000ApJ...538...92R, Clowe...2006ApJ...648L.109C, Shin...2022ApJ...934...43S}.
    Thus, a comparison of the kinematics and spatial distribution of galaxies and ICM can provide an important hint for the formation and evolution history of galaxy clusters (e.g., \citealp{Song...2017ApJ...842...88S, Song...2018ApJ...869..124S}).
    
    Among many clusters that we can study both gas and galaxies, we focus on the Perseus cluster (Abell 426; A426) in this study which has abundant X-ray data sets.
    This cluster is one of the nearest clusters ($z=0.017$) and is the brightest cluster in the X-ray sky.
    X-ray is one way to study ICM, as it has high temperatures ($\sim$keV) and emits X-ray via bremsstrahlung.
    Previous research on the ICM of the Perseus cluster focused on the structures and dynamics of gas in the central region.
    \citet{Fabian...2000MNRAS.318L..65F} investigated the structures in the central region of the Perseus cluster with the \textit{Chandra} data.
    \citet{Simionescu...2012ApJ...757..182S} compared the X-ray images from \textit{Chandra}, \textit{XMM-Newton}, and \textit{Suzaku} with numerical simulations and noted that the spiral features of the gas may be related to a past minor merger.
    \citet{Tamura...2014ApJ...782...38T} measured the bulk motion of ICM in the core region ($R \lesssim 5\arcmin$) with \textit{Suzaku}.
    They found a velocity structure west of the cluster, where an excess of the X-ray intensity exists.
    \citet{Hitomi...2016Natur.535..117H} presented high-resolution spectroscopic measurements of the gas bulk velocity and velocity dispersion in the central region ($\sim$100kpc) of the Perseus cluster.
    \citet{Sanders...2020A&A...633A..42S} analyzed the photometric and spectroscopic data from the \textit{XMM-Newton} archive.
    They found a correlation between the ICM velocity and temperature, especially giving attention to the bulk motion of gas in the eastern region of the cluster core.
    The X-ray data of the Perseus cluster have also been used for the search for dark matter \citep{Tamura...2015PASJ...67...23T, Tamura...2019PASJ...71...50T}.
    
    On the other hand, there are also active studies on the galaxy population of the Perseus cluster.
    For example, \citet{Aguerri...2020MNRAS.494.1681A} investigated the luminosity function of galaxies and dwarf galaxy populations with 963 newly measured redshifts.
    \citet{Meusinger...2020AA...640A..30M} presented a catalog of galaxies in the Perseus region and analyzed galaxy properties such as active galaxies and morphology.
    \citet{Khanday...2022MNRAS.515.5043K} used the Sloan Digital Sky Survey (SDSS; \citealp{York...2000AJ....120.1579Y}) data to examine the scaling relations of galaxies (e.g., the fundamental plane).
    However, there have been few studies focusing on the combined view of the galaxies and ICM in the Perseus cluster (e.g., \citealp{Tamura...2014ApJ...782...38T}).
    This is mainly because of the lack of deep spectroscopic data for the galaxies in the central region where we can make direct comparisons between the two components. 
    We have therefore conducted a deep spectroscopic survey for this cluster and made the comparison between galaxies and ICM.

    The aim of this paper is three-fold.
    First, we present a catalog of galaxies with measured redshifts in the central region of the Perseus cluster.
    Second, we introduce CausticSNUpy\footnote{https://github.com/woodykang/CausticSNUpy}, a Python implementation of the caustic technique for cluster membership determination.
    Third, we analyze the physical characteristics of the Perseus cluster, especially by comparing the spatial distribution and kinematics of galaxies with the X-ray-emitting ICM.   
    
    In Section \ref{sec:data}, we describe the data used in this study.
    We elaborate on the membership determination and examine the physical properties of the Perseus cluster in Section \ref{sec:res}.
    We discuss and conclude our study in Section \ref{sec:discuss_conclusion}.
    Throughout the paper, we adopt the standard $\Lambda$ cold dark matter ($\Lambda$CDM) cosmology with $\Omega_{m}=0.3$, $\Omega_{\Lambda}=0.7$, and $H_{0} = 100h\ \mathrm{km\ s^{-1}\ Mpc^{-1}}$.

\section{Data}          \label{sec:data}

    \subsection{Photometric Data}   \label{sec:data:phot}

        \begin{figure}
            \centering
            \includegraphics[width=1.0\linewidth]{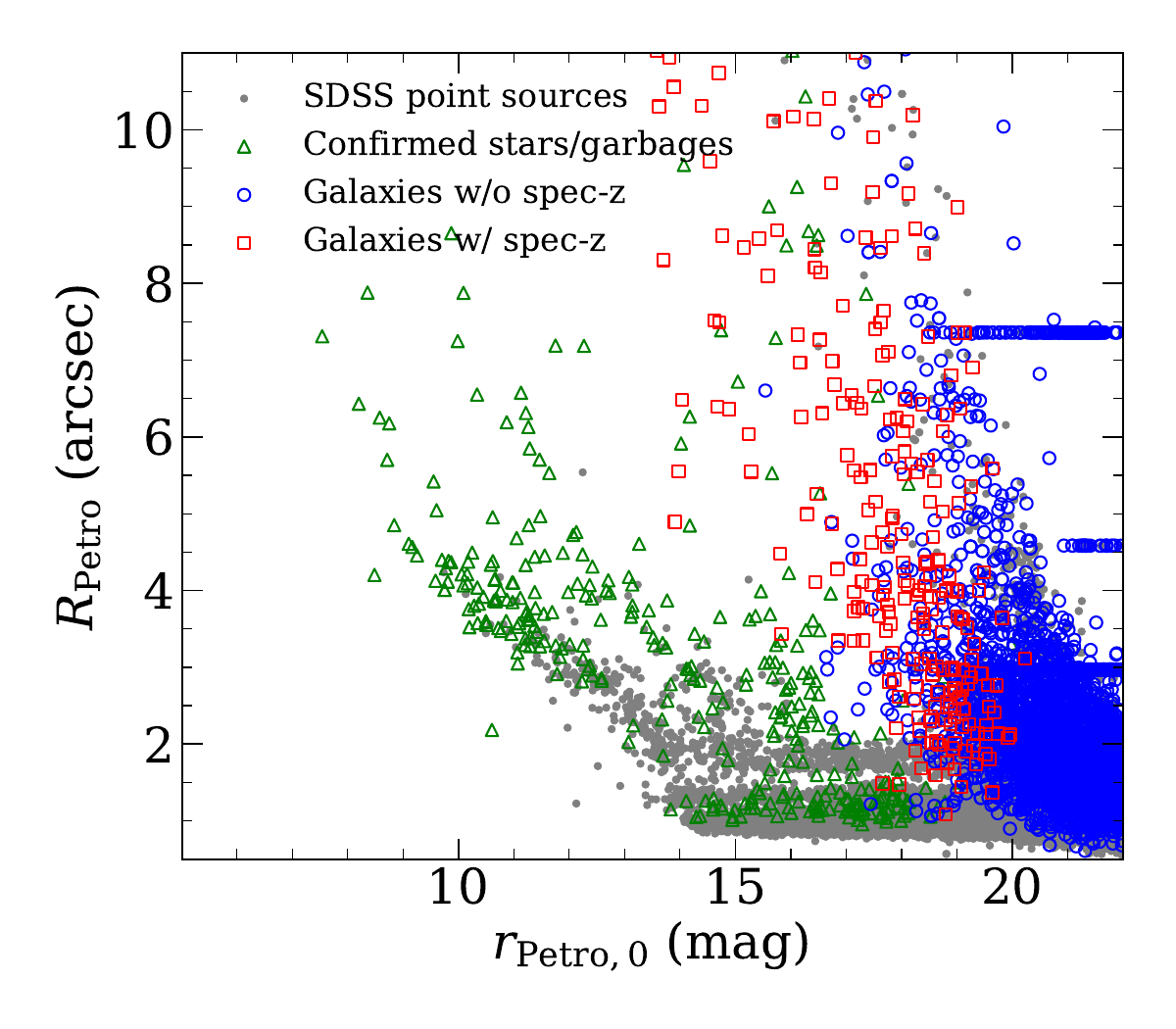}
            \caption{The size-magnitude diagram of objects in the SDSS photometric catalog. Gray dots represent point sources identified by SDSS, and green triangles are objects originally labeled as extended sources but visually confirmed to be point sources or artifacts. Blue circles and red squares are galaxies without and with spectroscopic redshifts. Only 20\% of the data points except the green triangles are plotted for clarity.
            \label{fig:1}
            }
        \end{figure}
        
        For the photometric data, we used the SDSS Data Release 17 (DR17; \citealp{Abdurrouf...2022ApJS..259...35A}).
        From this photometric catalog, we selected target galaxies for observation with MMT/Hectospec.
        Our goal for the observation was to obtain complete and uniform spectroscopic data in the Perseus cluster region.
        Therefore, we prioritized galaxies according to their \textit{r}-band apparent magnitude without any other criteria such as color selection.
        The SDSS photometric catalog provides a flag (\texttt{p\_probpsf}) that indicates if the object is likely to be a point source.
        However, some objects labeled as extended sources turned out to be images of saturated stars or artifacts (e.g., spikes). To exclude such objects from the target list, we conducted a visual inspection of all the targets plus any objects brighter than 17 mag in $r$-band were also inspected. A total of 745 of 3331 objects were relabeled as a result.
        In Figure \ref{fig:1}, we plot the size-magnitude plot of the objects in the SDSS photometric catalog.  We use the Petrosian radius $R_{\rm{Petro}}$ for the size and the extinction-corrected $r$-band magnitude $r_{\rm{Petro},0}$ for the magnitude. The green triangles are objects previously flagged as extended sources, but excluded after visual inspection. The plot confirms that the excluded objects are not likely to be extended objects.

    \subsection{Spectroscopic Data} \label{sec:data:spec}    

        \begin{deluxetable*}{ccccccc}
                \label{tab:field}
                \tablehead{
                    \colhead{Field ID} & \colhead{R.A.} & \colhead{Decl.} & \colhead{Observation Date} & \colhead{Exposure}  & \colhead{Number of} & \colhead{Number of}\\   
                                       & \colhead{(deg, J2000)} & \colhead{(deg, J2000)} & & \colhead{(minutes)} & \colhead{Targets} & \colhead{Measured Redshifts}
                }
                \caption{Observation with MMT/Hectospec}
                \startdata
                a426c\_2    &  49.965949  &  41.511704  & Nov 19 2014  &  $3\times20$  &  258  &  232 \\
                a426c21\_2  &  49.545053  &  41.179996  &  Dec 7 2021  &  $3\times20$  &  259  &  214 \\
                a426c21\_1  &  50.452605  &  41.179409  &  Dec 8 2021  &  $3\times20$  &  258  &  246 \\
                a426c21\_3  &  49.573721  &  41.940983  &  Dec 9 2021  &  $3\times20$  &  254  &  233 \\
                a426c21\_4  &  50.426230  &  41.941936  &  Dec 9 2021  &  $3\times20$  &  254  &  235
                \enddata
        \end{deluxetable*}

        \begin{figure*}
            \centering
            \includegraphics[width=0.85\linewidth]{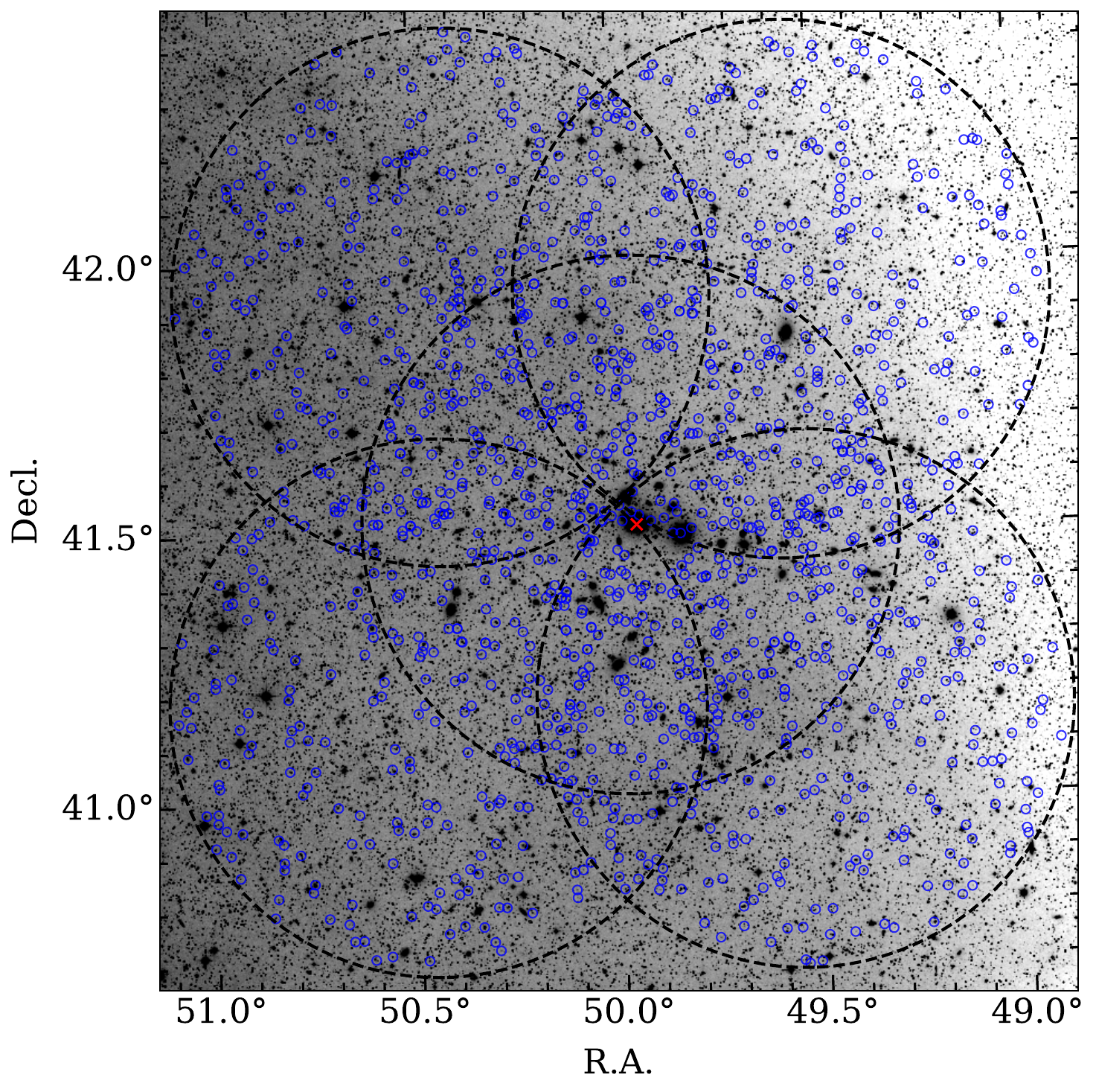}
            \caption{Observation fields of MMT/Hectospec in the Perseus cluster region.
                     Dashed circles indicate the five fields and open circles represent the positions of the observed galaxies.
                     The red cross shows the center of the cluster.
                     The Digitized Sky Survey image is shown in the background.
                     }
            \label{fig:field}
        \end{figure*}

        \begin{deluxetable*}{ccccccccc}
            \caption{Catalog of objects with redshift measured within 60\arcmin\ from the Perseus cluster center}
            \label{tab:catalog}
            \tablehead{
                \colhead{ID} & \colhead{SDSS DR17 ObjID\tablenotemark{a}} & \colhead{R.A.}  & \colhead{Decl.} & \colhead{$r_{\rm{Petro,0}}$} & \colhead{Flag\tablenotemark{b}} & \colhead{z\tablenotemark{c}} & \colhead{z Source\tablenotemark{d}} & \colhead{Member\tablenotemark{e}} \\
                             &                           & \colhead{(deg, J2000)} & \colhead{(deg, J2000)} & \colhead{(mag)}                  &                &             &                    &
            }
            \startdata
                   1 & 1237661121850770550 & 48.627453 & 41.415013 & 18.485&  1 &  $ -0.000259 \pm 0.000012 $ & 2 &0 \\
                   2 & 1237661083198882927 & 48.632691 & 41.555350 & 18.478&  1 &  $ -0.000057 \pm 0.000011 $ & 2 &0 \\
                   3 & 1237661121850835912 & 48.646550 & 41.309490 & 20.069&  1 &  $ -0.000717 \pm 0.000037 $ & 2 &0 \\
                   4 & 1237661121850835851 & 48.656534 & 41.362330 & 18.247&  1 &  $ -0.000049 \pm 0.000017 $ & 2 &0 \\
                   5 & 1237661083198882258 & 48.657297 & 41.532471 & 15.817&  1 &  $ -0.000272 \pm 0.000008 $ & 2 &0 \\
                   6 & 1237661121850835214 & 48.657786 & 41.343411 & 16.721&  0 &  $ -0.000228 \pm 0.000010 $ & 2 &0 \\
                   7 & 1237661121850835724 & 48.669263 & 41.436923 & 16.166&  0 &  $ 0.134324  \pm 0.000016 $ & 2 &0 \\
                   8 & 1237661121850835439 & 48.681708 & 41.275673 & 17.159&  1 &  $ -0.000257 \pm 0.000011 $ & 2 &0 \\
                   9 & 1237661083198882246 & 48.683231 & 41.565523 & 17.622&  0 &  $ -0.000038 \pm 0.000013 $ & 2 &0 \\
                  10 & 1237661055281397988 & 48.683388 & 41.823938 & 16.483&  0 &  $ -0.000091 \pm 0.000009 $ & 2 &0 
            \enddata
            \tablecomments{This table is available in its entirety in machine-readable form.}
            \tablenotetext{a}{ObjIDs of objects not found in DR17 are DR7 ObjIDs.}
            \tablenotetext{b}{(0) Extended source, (1) point source}
            \tablenotetext{c}{If redshift errors of galaxies from NED are not available, we set them to 0.0001.}
            \tablenotetext{d}{
            (1) This study,
            (2) SDSS,
            (3) \citet{Bilicki...2014ApJS..210....9B},
            (4) \citet{vandenBosch...2015ApJS..218...10V},
            (5) \citet{Brunzendorf1999},
            (6) \citet{Falco...1999PASP..111..438F}
            (7) \citet{Gandhi...2004MNRAS.348..529G},
            (8) \citet{Huchra...1999ApJS..121..287H},
            (9) \citet{Jorgensen...2018AJ....156..224J},
            (10), \citet{Miller...2001ApJS..134..355M},
            (11) \citet{Paturel2003},
            (12) \citet{Penny...2008...MNRAS.383..247P},
            (13) \citet{Penny...2014MNRAS...443...3381P},
            (14) \citet{Poulain1992},
            (15) \citet{Smith...2000MNRAS.313..469S},
            (16) \citet{Springob...2005ApJS..160..149S},
            (17) \citet{Zaw...2019ApJ...872..134Z},
            (18) \citet{Meusinger...2020AA...640A..30M}
            }
            \tablenotetext{e}{(0) Non-member, (1) Member of the Perseus cluster}
        \end{deluxetable*}
   
        We observed 1283 galaxies in the central region of the Perseus cluster with Hectospec on MMT 6.5m telescope \citep{Fabricant...2005PASP..117.1411F}, as summarized in Table \ref{tab:field}.
        Hectospec is a multiobject spectrograph with 300 fibers whose field of view is 1\arcdeg\ in diameter.
        We observed a total of five fields with MMT/Hectospec, covering a region of approximately 1\arcdeg\ in radius.
        The observed fields are shown in Figure \ref{fig:field}.
        After the observation, we reduced the data with HSRED v2.0, an IDL pipeline provided by the MMT Observatory.
        With the resulting 1D spectra, we measured the redshifts of the galaxies using RVSNUpy (Kim et al. in preparation).
        RVSNUpy is a Python package that measures the redshift of a spectrum by cross-correlating the given spectrum with various spectral templates.
        Along with the redshift value for each spectrum, the package yields the \citet{Tonry...1979AJ.....84.1511T} $r$ value ($r_{TD}$) which can be used as a measure of reliability.
        \citet{Geller...2014ApJS..213...35G, Geller...2016ApJS..224...11G} confirmed by visual inspection that redshifts with $r_{TD}>4$ are reliable.
        We used galaxies whose $r_{TD}$ are greater than 4.5 to be conservative.
        The sum of the number of reliable redshifts in each field is 1160, as can be inferred from Table \ref{tab:field}.
        Among the galaxies with reliable redshifts, 9 were observed twice and have two redshift values.
        For these galaxies, we use the redshifts with higher $r_{TD}$.
        This results in 1151 galaxies with reliable redshifts, among which 1148 galaxies are within 60\arcmin\ from the center of the cluster.
        We also incorporated redshift data from the literature.
        We retrieved 502 redshifts from SDSS DR17, 35 from NASA/IPAC Extragalactic Database (NED), and 9 from \citet{Meusinger...2020AA...640A..30M}; among those, 265 galaxies (SDSS), 29 galaxies (NED), and 5 galaxies \citep{Meusinger...2020AA...640A..30M} are within 60\arcmin.
        Thus, the total number of galaxies with measured redshifts at $R<60\arcmin$ is 1447.
        It should be noted that there was an extensive galaxy redshift survey on this cluster by \citet{Aguerri...2020MNRAS.494.1681A} who measured 963 redshifts of the galaxies. However, we could not include them in this study because their data are not publicly available. We omit them in the evaluation of the usefulness of our MMT/Hectospec data in the following paragraphs. We note that they covered the cluster region with $R < 147\arcmin$. However, there is only one pointing out of 21 pointings covering $R < 30\arcmin$, suggesting that the overlap of the data between the two studies is not significant.

        \begin{figure*}
            \centering
            \includegraphics[width=0.8\linewidth]{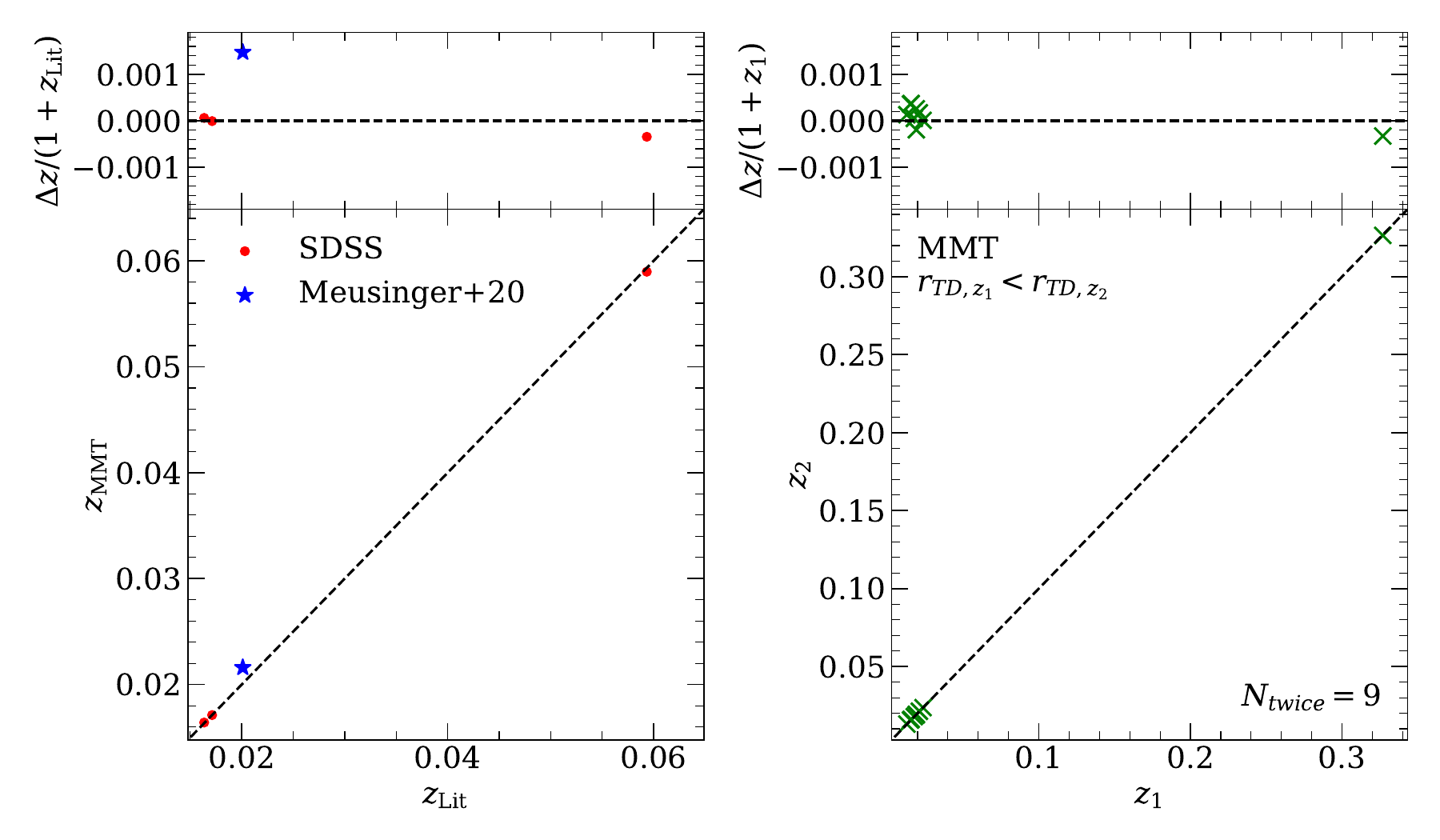}
            \caption{
            Comparison of redshifts between sources (left) and within MMT/Hectospec observation in this study (right). On the left panels, $z_{\rm{MMT}}$ and $z_{\rm{Lit}}$ denote redshifts measured by MMT/Hectospec and those from the literature, respectively. On the right panels, $z_{2}$ are redshifts with higher $r_{TD}$ than $z_{1}$.
            }            
            \label{fig:z_comp}
        \end{figure*}

        We compare the redshift measurements of galaxies in Figure \ref{fig:z_comp}.
        The left panels show the comparison between redshifts in the literature ($z_{\rm{Lit}}$) and the redshift measured by MMT/Hectospec in this study ($z_{\rm{MMT}}$).
        There are three overlapping galaxies from SDSS (red circle) and one from \citet{Meusinger...2020AA...640A..30M} (blue star).
        The plot shows a general agreement between the two.
        The right panels show the comparison of redshifts of the 9 galaxies measured twice with MMT/Hectospec in this study.
        The plot again shows that the repetitive measurements agree well.
        We denote $z_{2}$ as the redshift with a higher $r_{TD}$ and $z_{1}$ as the other.
        
        We constructed a catalog containing the SDSS ObjID, coordinates, extinction-corrected \textit{r}-band Petrosian magnitudes $r_{\rm{Petro}, 0}$, measured redshifts, and cluster membership (see Section \ref{sec:res:member} for membership determination).
        The catalog is given in Table \ref{tab:catalog}.
        For the full presentation of the data, we include in the catalog three galaxies whose spectra are measured by MMT/Hectospec but outside 60\arcmin\ from the cluster center.

        \begin{figure*}
            \centering
            \includegraphics[width=0.7\linewidth]{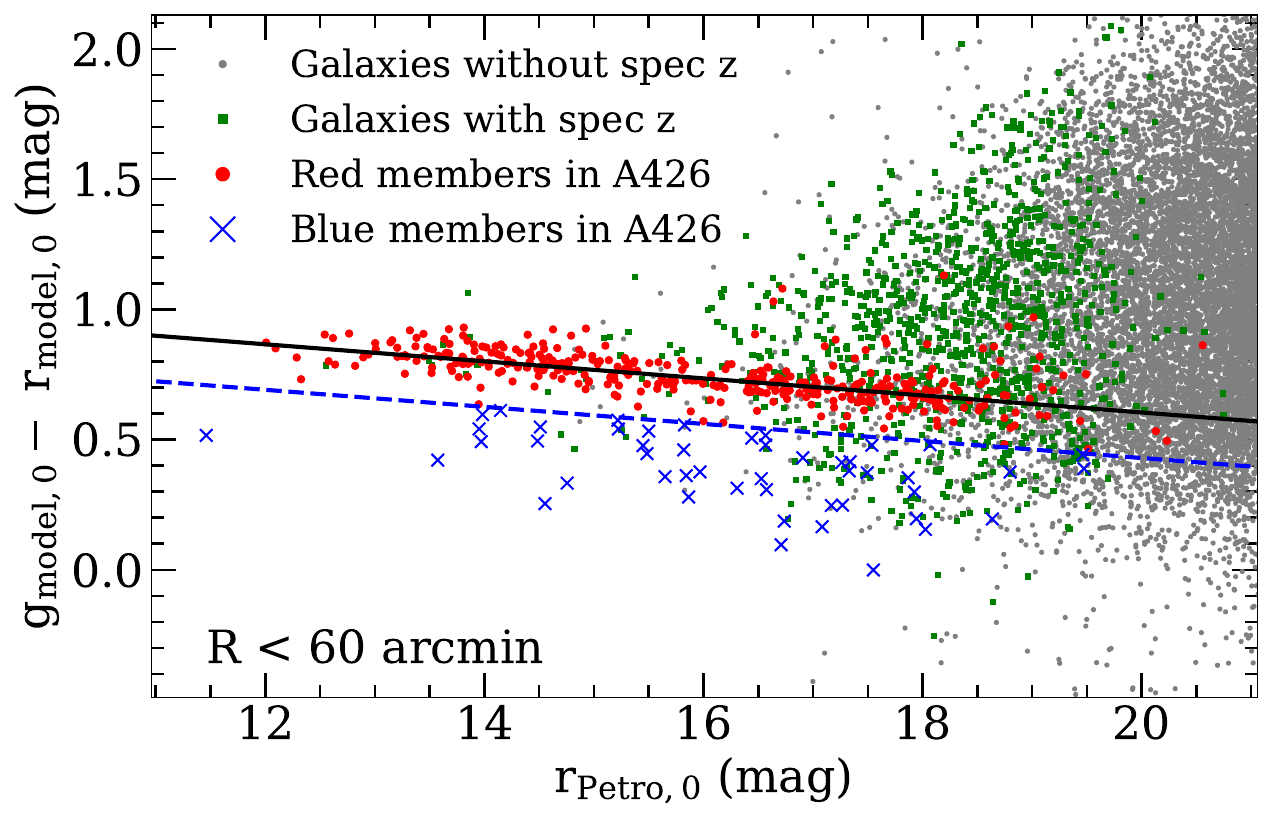}
            \caption{Color-magnitude diagram of galaxies within 60\arcmin\ from the cluster center. Gray dot, green square, red circle, and blue cross indicate galaxies without spectroscopic redshifts, galaxies with spectroscopic redshifts but interlopers, red members of the cluster, and blue members of the cluster, respectively. The black solid line represents the red sequence calculated by linear fitting. The blue dashed line is $3\sigma$ bluer than the red sequence, which is the criterion for classifying blue and red members.}
            \label{fig:cmd}
        \end{figure*}
        
        Figure \ref{fig:cmd} shows the color-magnitude diagram (CMD) of galaxies within 60\arcmin\ from the cluster center.
        The cluster members form a clear red sequence in the diagram.
        To partition the cluster members into blue and red members, we follow the method in \citet{Hwang...2014ApJ...797..106H}.
        We first draw the red sequence by linear fitting member galaxies with $r_{\rm{Petro}, 0}<18$ and $g_{\rm{model},0} - r_{\rm{model}, 0} > 0.5$.
        The best-fit line (black solid line) is $g_{\rm{model},0} - r_{\rm{model}, 0} = -0.033 r_{\rm{Petro}, 0} + 1.26$ with scatter $\sigma = 0.058$.
        Galaxies bluer than $3\sigma$ from the red sequence are defined as blue members, and the others as red members.
        The $3\sigma$ boundary is shown in the blue dashed line in Figure \ref{fig:cmd}.
        Out of 418 member galaxies within 60\arcmin, 370 are classified as red members and 48 as blue members.
        While inspecting the CMD, we found that four cluster members had unusual apparent magnitudes on the CMD.
        Those objects have adjacent foreground stars, which probably affected the magnitude measurements.
        For the four galaxies, we replaced the Petrosian magnitudes with the fiber magnitudes.
        
\begin{figure*}
            \centering
            \includegraphics[width=0.9\linewidth]{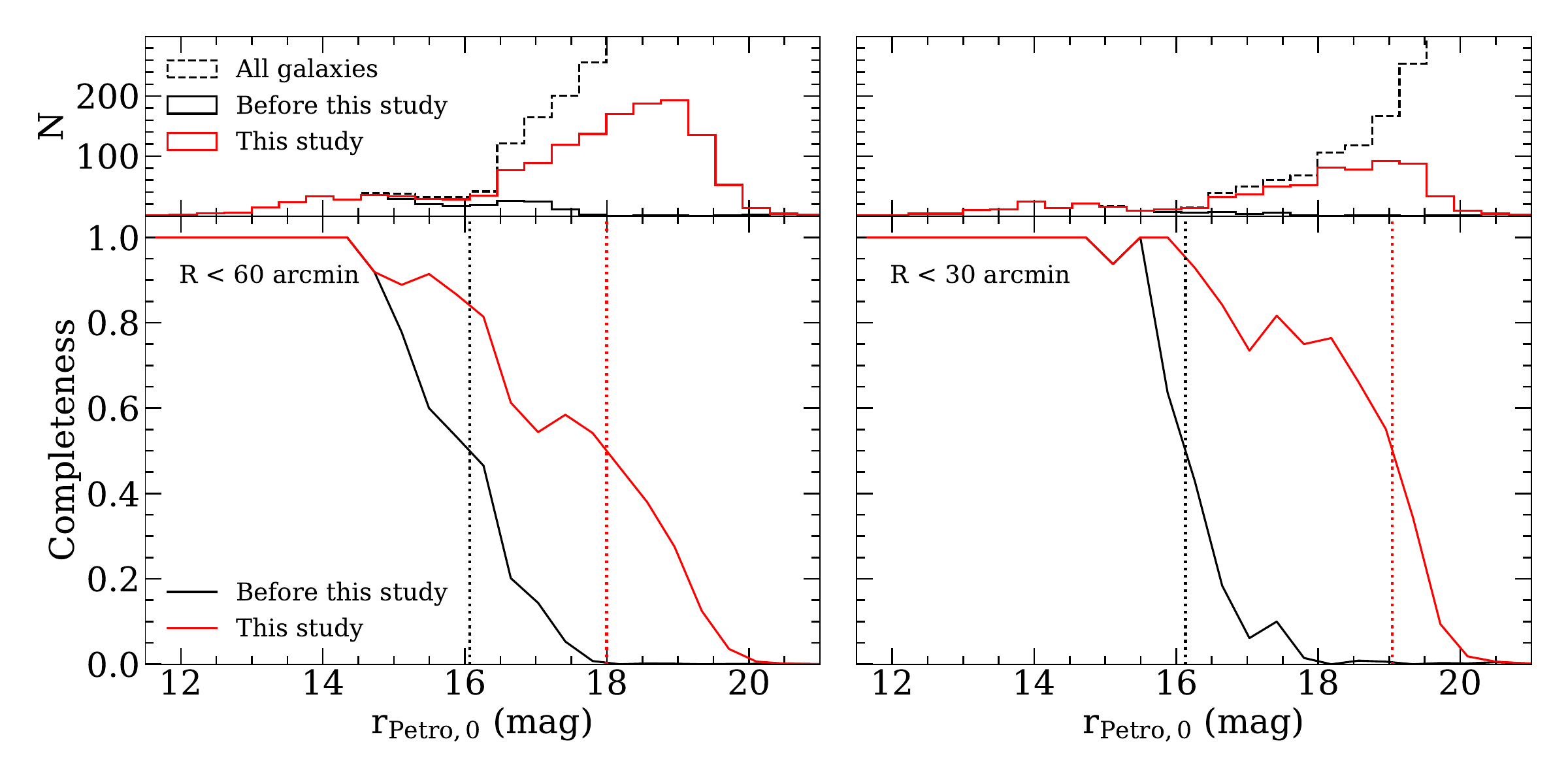}
            \caption{Top left: histogram of galaxies within 60\arcmin\ from cluster center for all galaxies (black dashed line), galaxies with redshifts before this study (solid black line), and galaxies with redshifts from this study (solid red line). Bottom left: differential spectroscopic completeness as a function of magnitude for data before this study (solid black line) and from this study (solid red line). Black and red dotted vertical lines show the magnitudes corresponding to the differential completeness of 50\%. The magnitudes are 16.1 mag and 18.0 mag, respectively. Top right and bottom right: Same as the left panels, but for galaxies within 30\arcmin. The magnitudes at which the completeness is 50\% are 16.1 mag for data before this study and 19.1 mag for data from this study.}
            \label{fig:completeness}
        \end{figure*}

        Figure \ref{fig:completeness} shows the comparison of the spectroscopic completeness before this study and with the new MMT/Hectospec data added.
        We plot the spectroscopic completeness for two regions: $R<60'$ (left) and $R<30'$ (right).
        The top panels show the histograms of all galaxies (black dashed lines), galaxies with spectroscopic redshifts before this study (solid black lines), and galaxies with redshifts obtained in this study (solid red lines), as functions of $r_{\rm{Petro}, 0}$.
        The bottom panels show the differential completeness as functions of $r_{\rm{Petro}, 0}$ before this study (solid black lines) and with data used in this study (solid red lines).
        The black and red vertical dotted lines indicate the magnitudes at which the differential completeness drops below 50\%. The 50\% magnitude limit was pushed from 16.1 mag to 18.0 mag for galaxies within 60\arcmin, and from 16.1 mag to 19.1 mag for those within 30\arcmin.
        The cumulative spectroscopic completeness is 67\% for galaxies with $R < 60\arcmin$ and $r_{\mathrm{Petro,0}} \leq 18.0$, and 76\% for $R < 30\arcmin$ and $r_{\mathrm{Petro,0}} \leq 19.1$. 
        As a result, our new data play an essential role in increasing the spectroscopic completeness, especially for the $R<30\arcmin$ region.

\begin{figure*}
            \centering
            \includegraphics[width=1.0\linewidth]{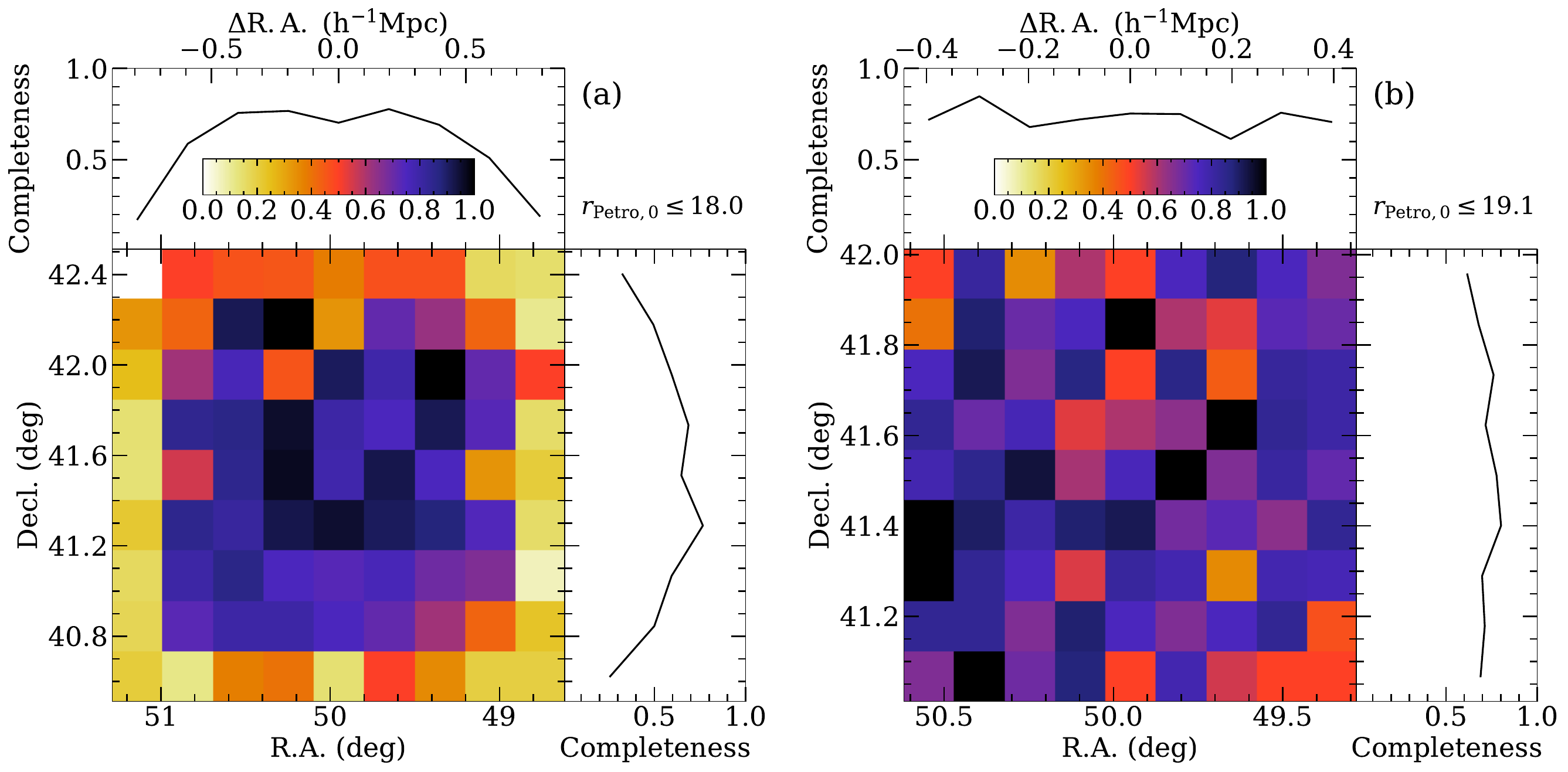}
            \caption{(a) Bottom left: Spectroscopic completeness of galaxies with $r_{\rm{Petro}, 0} \le 18.0$ as a function of right ascension and declination. The field size is $120\arcmin \times 120\arcmin$ with its center aligned to the center of the Perseus cluster. Marginal completeness in the right ascension (top) and declination (right) are also given. (b) Same as (a), but for the galaxies brighter than 19.1 mag in the $60\arcmin \times 60\arcmin$ field.}
            \label{fig:2dcompleteness}
        \end{figure*}

        To confirm the spatial uniformity of the completeness, Figure \ref{fig:2dcompleteness} shows the 2D spectroscopic completeness.
        The fields of view for (a) and (b) of Figure \ref{fig:2dcompleteness} are $120\arcmin \times 120\arcmin$ and $60\arcmin \times 60\arcmin$, corresponding to $R<60\arcmin$ and $R<30\arcmin$, respectively.
        The bottom left panels of (a) and (b) show the 2D spectroscopic completeness with the centers of the field of view aligned to the cluster center.
        The top panels and the right panels show the marginal completeness along R.A. and Decl., respectively.
        The completeness is calculated for the galaxies brighter than 18.0 mag for (a) and 19.1 mag for (b), which are the 50\% magnitude limit as described in the preceding paragraph.
        The completeness for $R<60\arcmin$ is high in the central region but drops rapidly at the edges.
        On the other hand, albeit with the fainter magnitude limit, the $R<30'$ region maintains overall uniform and high completeness.
        The reason for the higher completeness in the inner region is due to the overlapping observation fields in the central region as shown in Figure \ref{fig:field}.
        This is consistent with the results from Figure \ref{fig:completeness}.
        For this reason, we use the $R<30\arcmin$ region with a magnitude limit of 19.1 for analyses where spatial distribution is considered important.

        \begin{figure}
            \centering
            \includegraphics[width=0.9\columnwidth]{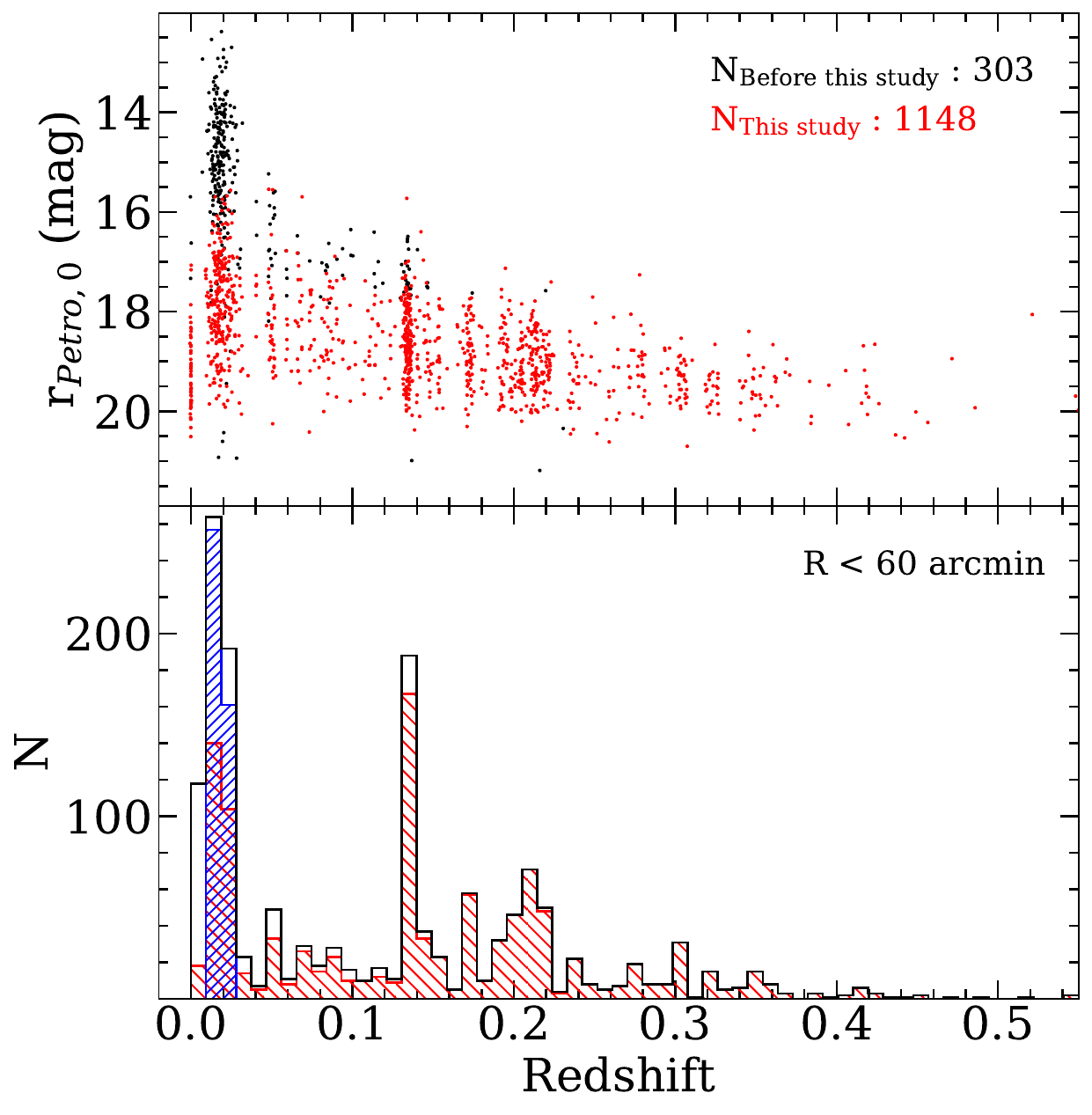}
            \caption{(Top) Distribution of $r_{\rm{Petro}, 0}$ along redshift. Black dots represent individual galaxy redshifts before this study, while red dots represent galaxies observed with MMT/Hectospec. (Bottom) Black and red lines show the histogram of the whole galaxies with spectroscopic redshifts and galaxies whose redshifts are measured with MMT/Hectospec in the $R < 60\arcmin$ region, respectively. The blue histogram shows the galaxies determined as cluster members.}
            \label{fig:new_redshift}
        \end{figure}
        
        Figure \ref{fig:new_redshift} shows the comparison between the redshift distribution of the data available before this study and the data used in this study. The top panel shows the apparent magnitudes of galaxies as a function of redshift. Black dots indicate galaxy data before this study, and red dots indicate those observed with MMT/Hectospec. The bottom panel shows the number of galaxies with measured redshifts. The black line represents all galaxies, consisting of those from MMT/Hectospec, SDSS, and NED. The red hatched histogram shows only the galaxies observed with MMT/Hectospec. The blue histogram represents the member galaxies of the Perseus cluster. As expected, the member galaxies are concentrated around the redshift of the cluster center ($z=0.01756$). As Figure \ref{fig:new_redshift} proves, the MMT/Hectospec data complement the existing data set with faint galaxies of relatively higher redshifts.
    
\section{Results}    \label{sec:res}
        \subsection{Cluster Membership Determination}    \label{sec:res:member}
        
\begin{figure*}
            \centering
            \includegraphics[width=0.8\linewidth]{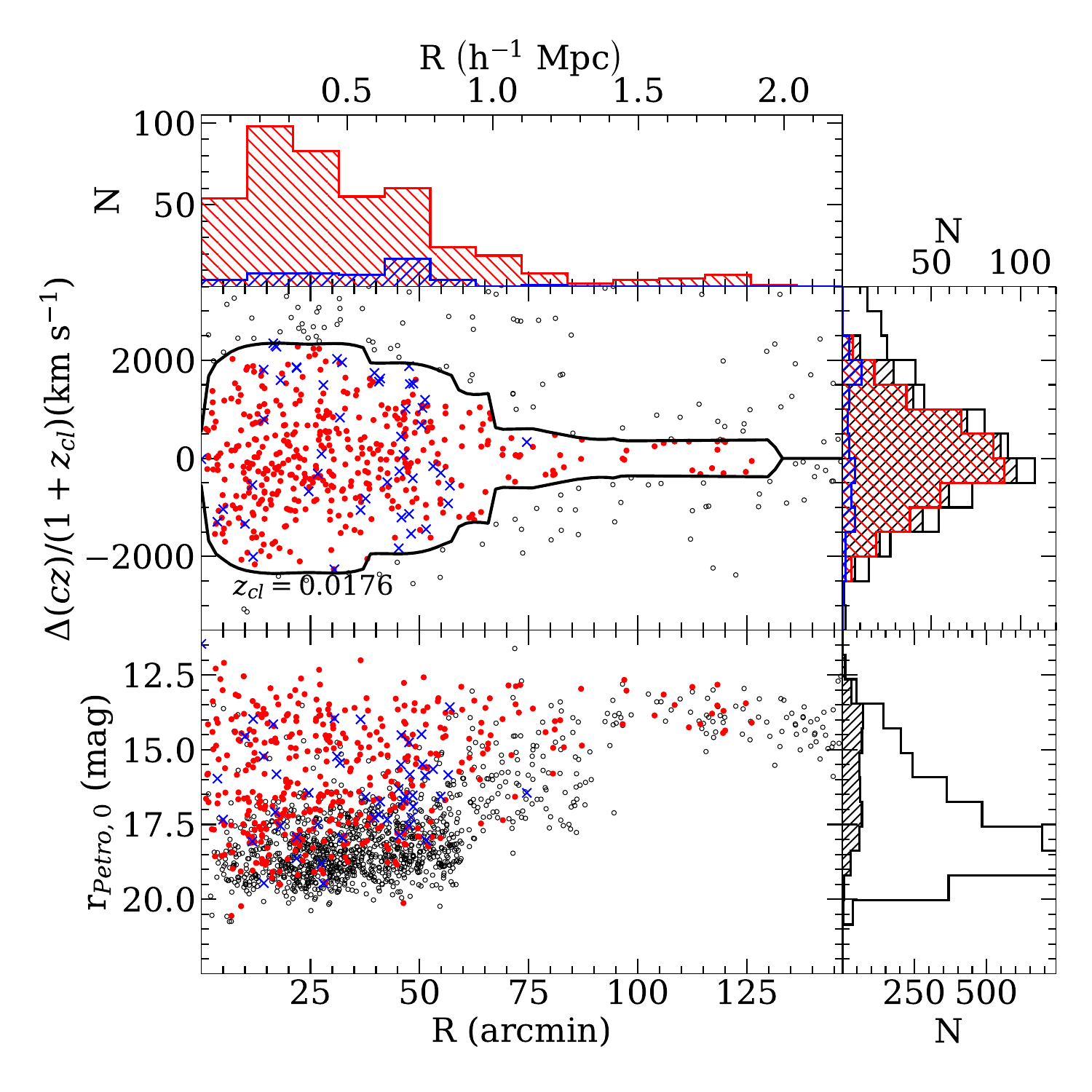}
            \caption{
                The caustics are plotted in the middle left panel, along with red member galaxies (red circles), blue member galaxies (blue crosses), and non-member galaxies (black open circles).
                The bottom left panel shows the distribution of apparent magnitude as a function of clustercentric distance.
                Histograms of galaxies in the field of the Perseus cluster are represented with respect to clustercentric distance (top left), radial velocity (middle right), and apparent magnitude (bottom right).
                Red and blue histograms are for red and blue members, respectively.
                Black histograms are for all member galaxies, while open histograms are for all observed galaxies.
                }
                \label{fig:rvmag}
        \end{figure*}

        To analyze the cluster property, the separation of cluster member galaxies from background and foreground galaxies is necessary. One method to determine cluster membership is to plot galaxies in the 2D pseudo phase space known as the redshift space (radial velocity v.s. projected clustercentric distance; middle left panel of Figure \ref{fig:rvmag}). Galaxies in a cluster form a trumpet-shaped distribution in the redshift space, whose envelopes are referred to as caustics \citep{Kaiser...1987MNRAS.227....1K, Regos...1989AJ.....98..755R, Diaferio...1997ApJ...481..633D}. The caustics are related to the escape velocity at a given radius and in turn, galaxies lying outside the caustics can be regarded as interlopers \citep{Diaferio...1999MNRAS.309..610D}. \citet{Serra...2013ApJ...768..116S} tested the caustic technique on galaxy clusters produced from N-body simulation clusters and reported that 95\% of the true cluster members were recovered and only 8\% of the interlopers were identified as members within $3r_{200}$, where $r_{200}$ is the radius at which the average density within a sphere is equal to 200 times the critical density of the universe. This confirms that the caustic technique is reliable for distinguishing cluster members from the back- and foreground galaxies.

        There are software implementations of the caustic technique. 
        For example, Ana Laura Serra \& Antonaldo Diaferio (2014) developed the Caustic App, a standalone software tool for applying the caustic method.
        The software was used in previous studies such as \citet{Song...2017ApJ...842...88S, Song...2018ApJ...869..124S}.
        Their tool is limited to a certain operating system (Linux) and does not support recent OS versions.
        On the other hand, a Python package called causticpy is also available.
        The package is based on \citet{Gifford...2013ApJ...773..116G} and \citet{Gifford...2013ApJ...768L..32G}.
        Different assumptions and implementation details are adopted in \citet{Gifford...2013ApJ...768L..32G} than in the traditional caustic technique, including the assumption of Navarro-Frenk-White profile and using static kernel density estimation.
        
        In this study, we develop CausticSNUpy, a Python code to run the caustic technique.
        The code follows the procedures elaborated in detail by \citet{Diaferio...1999MNRAS.309..610D}, \citet{Serra...2011MNRAS.412..800S}, and \citet{Serra...2013ApJ...768..116S}.
        The code first finds candidate members using a hierarchical clustering algorithm.
        The code then calculates the galaxy number density in the redshift space via adaptive kernel density estimation.
        The caustics are where the number density reaches a certain threshold.
        The final member galaxies are determined as those lying inside the caustics in the redshift space.
        In Appendix \ref{sec:app:caustic}, we compare the caustics of our code with that of the Caustic App.
        
        The caustic method also provides an estimate of the cluster center.
        The right ascension and declination of the center of the cluster are calculated as the peak of the galaxy number density on the sky, and the redshift as the median of the candidate member redshifts \citep{Serra...2011MNRAS.412..800S}.
        The center of the Perseus cluster calculated by the caustic method is $(03^{\rm h} 19^{\rm m} 21\fs70, 41\arcdeg29\arcmin42\farcs44)$ and $cz = 5006\mathrm{km\ s^{-1}}$.
        The brightest cluster galaxy (BCG) can also be used as the cluster center.
        For the Perseus cluster, the BGC is NGC 1275 whose coordinates are $(03^{\rm h} 19^{\rm m} 48\fs16, 41\arcdeg 30\arcmin 42\farcs1)$ and $cz = 5264 \mathrm{km\ s^{-1}}$ \citep{Hudson...2001MNRAS.327..265H}.
        The offset between the cluster center derived from the caustic technique and NGC 1275 is 5\farcm05 on the sky and $258 \rm{km\ s^{-1}}$ in redshift.
        We fix the cluster center to the coordinates of NGC 1275 for consistency with other studies.
        
        The middle left panel of Figure \ref{fig:rvmag} shows the result of the caustic technique run with CausticSNUpy.
        The black solid line represents the caustics. The red circles, blue crosses, and black open circles indicate red members, blue members, and non-members, respectively.
        A total of 468 galaxies were determined as cluster members out of 1699 galaxies with spectroscopic redshifts.
        Among them, 418 galaxies were identified as members within 60\arcmin\ and 246 galaxies within 30\arcmin. 
            
        The top left and middle right panels show the histograms of galaxies along the clustercentric distance and the radial velocity.
        Black, red, and blue hatched histograms represent the histograms of all members, red members, and blue members, respectively.
        The empty histogram is for all galaxies with redshifts. 
        The distribution of apparent magnitudes is shown in the bottom left panel.
        The magnitude limits for MMT/Hectospec (20.5 mag) and SDSS (17.7 mag) are apparent in $R < 60\arcmin$ and $R < 90\arcmin$, respectively.
            
        \subsection{Global Kinematics of the Cluster Traced by Galaxies}
            \subsubsection{Velocity Dispersion Profile} \label{sec:res:veldisp}

            \begin{figure*}
                \centering
                \includegraphics[width=0.65\linewidth]{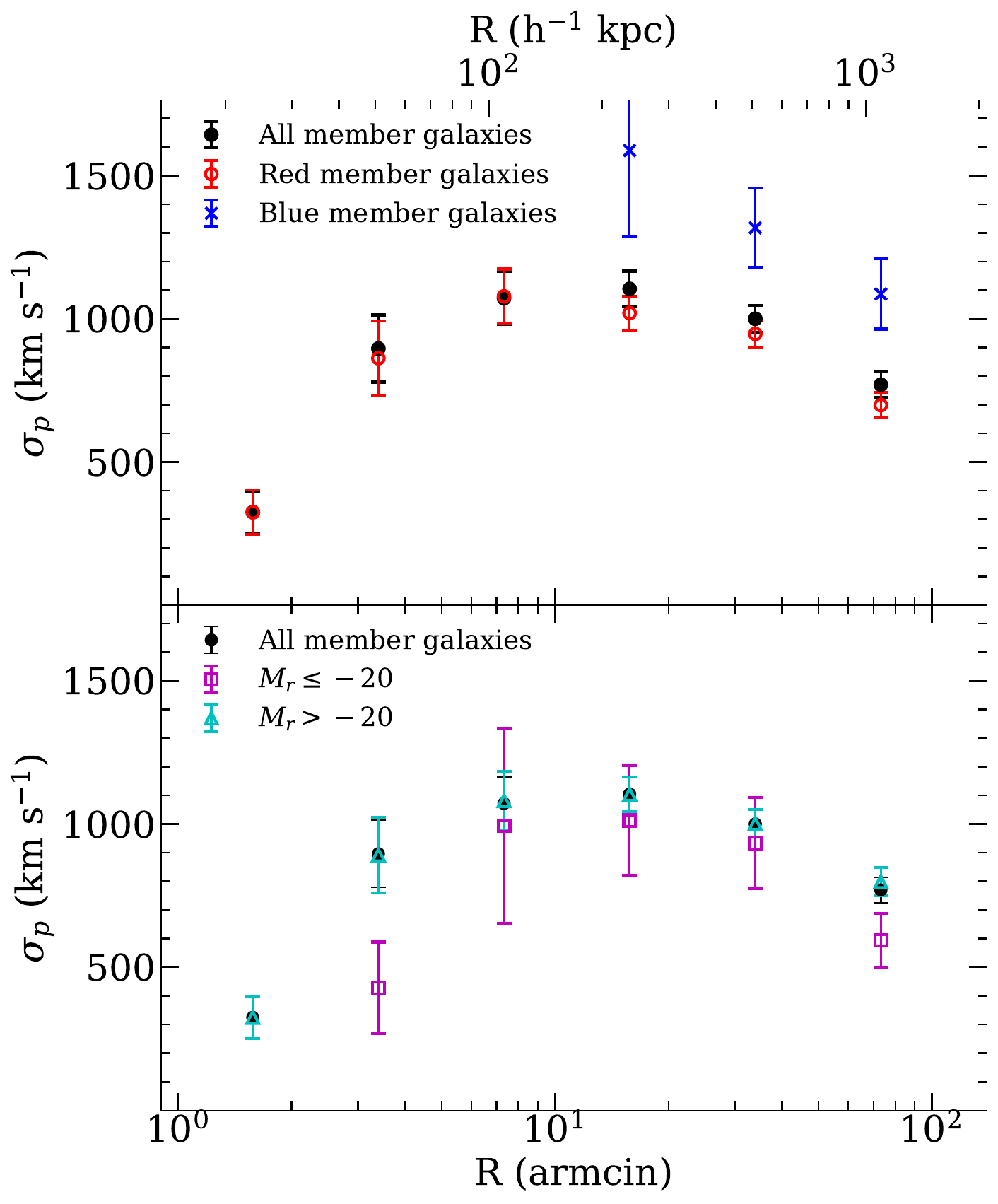}
                \caption{
                    Radial velocity dispersion profile of galaxies.
                    On both panels, the black circles represent the velocity dispersion of all member galaxies.
                    (Top) Galaxies are divided into red member galaxies (red open circles) and blue member galaxies (blue crosses).
                    (Bottom) Galaxies are divided by their absolute magnitude. Magenta squares represent bright member galaxies and cyan plus signs represent faint member galaxies. 
                    Only bins with three or more galaxies are used.
                    Error bars of galaxy velocity dispersion are resampling errors.
                }
                \label{fig:veldisp}
            \end{figure*}
            
            Here, we present the velocity dispersion profile of member galaxies in the Perseus cluster (Figure \ref{fig:veldisp}) and provide related scientific questions.
            In Figure \ref{fig:veldisp}, we divide the sample into subsets according to two criteria: color (top) and absolute magnitude (bottom).
            This is to investigate the dependence of velocity dispersion profile on cluster mass tracer properties.
            The colors of member galaxies are defined in Section \ref{sec:data:spec}, and the absolute magnitude is calculated by taking into account the K- and E-correction,
            \begin{equation}
                M_{r} = r_{Petro,0} - DM - K(z) + E(z)
            \end{equation}
            where $M_{r}$ is the $r$-band absolute magnitude, $DM=5\log (D_{L}/10\rm{pc})$ is the distance modulus, $D_{L}$ is the luminosity distance, $K(z)$ is the K-correction, and $E(z)$ is the luminosity evolution correction.
            For the K-correction, we use the equation presented by \citet{Choi...2007ApJ...658..884C}:
            \begin{equation}
            \begin{split}
                K(z) = & 3.0084(z-0.1)^{2} + 1.0543(z-0.1) \\
                       & - 2.5\log(1+0.1)
            \end{split}
            \end{equation}
            We calculate the luminosity evolution correction as $E(z)=1.6(z-1)$, given by \citet{Tegmark...2004ApJ...606..702T}.
            We choose $-20$ mag as the magnitude cut to divide galaxies, which is approximately the characteristic magnitude of a cluster luminosity function.
            As the absolute magnitude can used as a proxy for the stellar mass of a galaxy, we can regard brighter galaxies as more massive ones.
            
            Previous studies showed that red member galaxies are reliable tracers of the cluster mass \citep{Rines...2013ApJ...767...15R, Geller...2014ApJ...783...52G, Song...2017ApJ...842...88S}.
            Also, in a galaxy cluster, less massive galaxies have joined the system rather recently.
            Thus, low-mass member galaxies may show a velocity dispersion larger than that of the high-mass galaxies because of high velocity anisotropy \citep{Biviano...2004AA...424..779B, Hwang...2008ApJ...676..218H}.
            The top panel shows that the inclusion of blue members does not significantly affect the measured velocity dispersion profile.
            In the bottom panel, the velocity dispersion of bright galaxies shows a slight offset from the total profile due to the small number of galaxies satisfying $M_{r} \leq -20$.
            
            The velocity dispersion profile of cluster galaxies can provide useful information about the mass profile of the cluster or the velocity anisotropy of cluster galaxies.
            Fixing one of the two parameters allows the estimation of the other.
            For example, \citet{Hwang...2008ApJ...676..218H} constrained the velocity anisotropies of 10 clusters from the velocity dispersion profiles of galaxies and the mass profiles inferred from X-ray observation data (see also \citealp{Biviano...2004AA...424..779B}).
            On the other hand, studies such as \citet{Katgert...2004ApJ...600..657K} assumed the velocity anisotropies to calculate the mass profiles of clusters.
            Furthermore, \citet{Geller...2014ApJ...783...52G} suggested that the stellar velocity dispersion profile of the BCG could provide a complementary measure of the cluster potential.
            A similar analysis can be done for this cluster with the dispersion profile in this study, which we leave for future works.
                            
            \subsubsection{Detecting Signals of Rotation}   \label{sec:res:rot}
                \begin{figure*}
                    \centering
                    \includegraphics[width=0.85\linewidth]{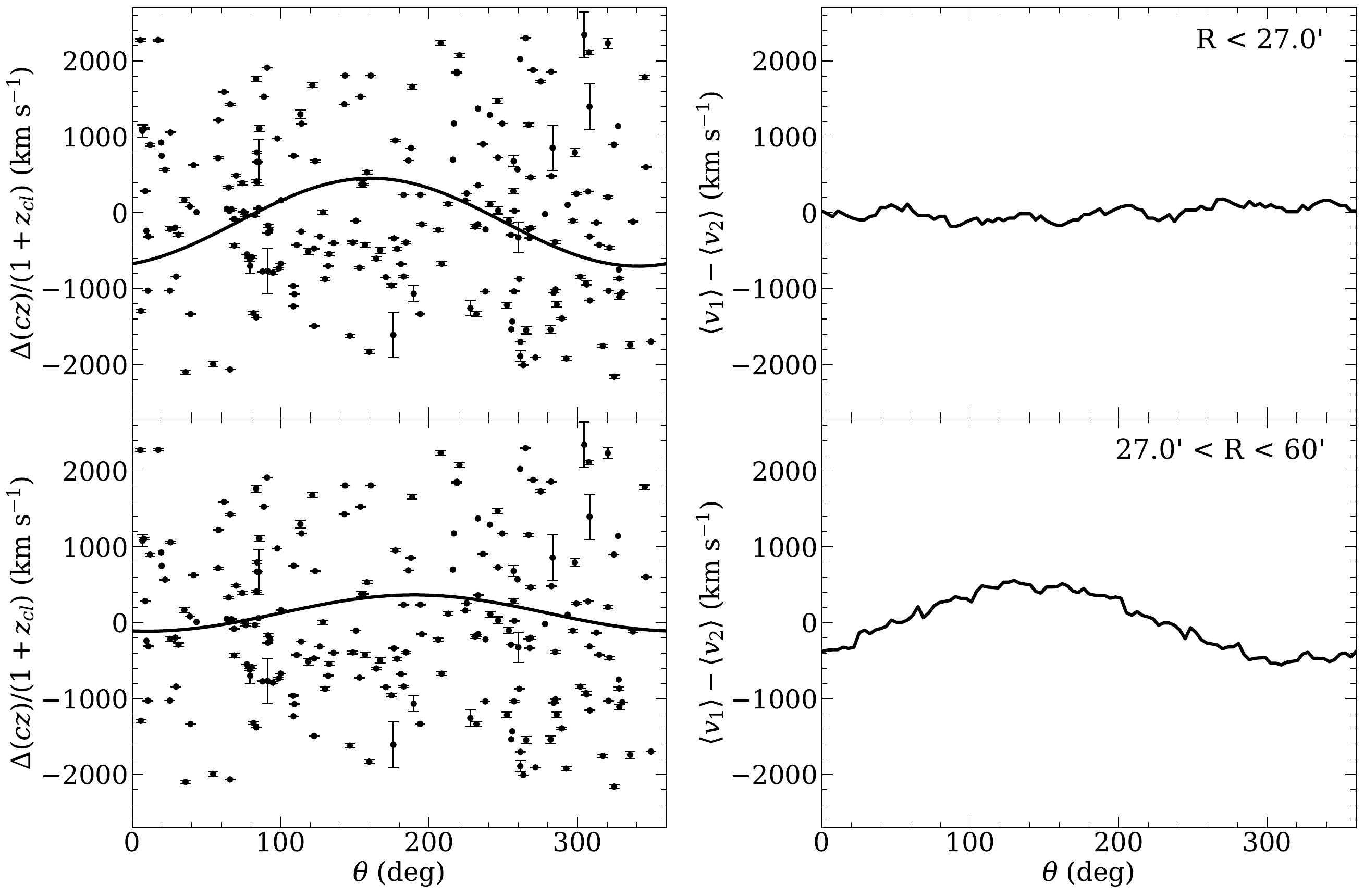}
                    \caption{
                    Rotation velocity of the Perseus cluster.
                    (Top left) Individual circles indicate the radial velocities of member galaxies within 27\arcmin\ as a function of position angle $\theta$.
                    The black line represents the fitted curve according to Equation \ref{eq:rot_Hwang}.
                    (Top right) Difference between the mean velocities $\langle v_{1} \rangle$, $\langle v_{2} \rangle$ of member galaxies within 27\arcmin\ in two regions divided by a line with position angle $\theta$.
                    (Bottom left and right) Same as the top panels, for galaxies in $27\arcmin < R < 60\arcmin$.
                    }
                    \label{fig:rotation}
                \end{figure*}

                \begin{figure}
                    \centering
                    \includegraphics[width=1.0\columnwidth]{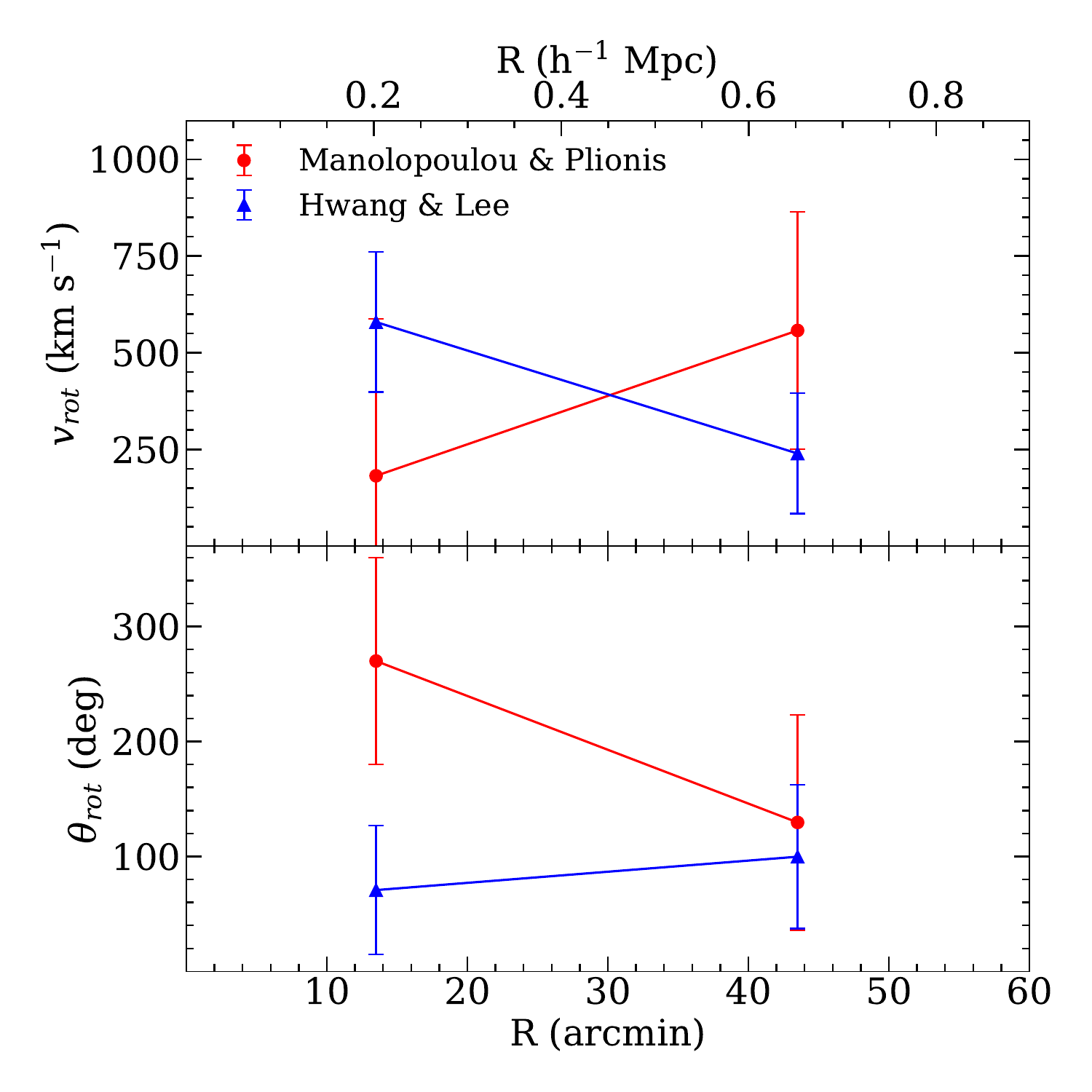}
                    \caption{
                        Rotation profile of the Perseus cluster.
                        (Top) Rotation amplitude.
                        (Bottom) Position angle of the rotation axis. Values in blue triangles are measured with \citet{Hwang...2007ApJ...662..236H} method and those in red circles with \citet{Manolopoulou...2017MNRAS.465.2616M}. Error bars are resampling errors.
                        The border between the inner region and the outer region is set to 27\arcmin\ so that there is almost the same number of galaxies in each region.
                    }
                    \label{fig:rotation_profile}
                \end{figure}
            
                A cluster can be rotating, by obtaining angular momentum from mergers with other clusters or due to lasting initial angular momentum since the formation of the cluster.
                There were several methods to detect and quantify the global rotation of a cluster.
                For example, \citet{Hwang...2007ApJ...662..236H} fit the radial velocities of galaxies with a sinusoidal function with regard to the position angle on the sky.
                The fitting function is 
                \begin{equation} \label{eq:rot_Hwang}
                    v_{p} ( \theta ) = v_{sys} + v_{rot} \sin (\theta - \theta_{rot})
                \end{equation}
                where $\theta$ is the position angle of each galaxy measured from north to east, $\theta_{rot}$ is the projected rotation axis on the sky, $v_{p}$ is the observed radial velocity, $v_{sys}$ is the systematic velocity of the cluster, and $v_{rot}$ is the rotation amplitude.
                The observed quantities are $v_{p}$ and $\theta$ and $v_{sys}$, $v_{rot}$, and $\theta_{rot}$ are fitting parameters.
                On the other hand, \citet{Manolopoulou...2017MNRAS.465.2616M} proposed another idea.
                Their algorithm first partitions the galaxies into two regions divided by a line projected on the sky and passing through the cluster center.
                The mean velocities of galaxies in the two regions, $\langle v_{1} \rangle $ and $\langle v_{2} \rangle$, are calculated.
                Then, the algorithm rotates the divider from north to east and finds when the difference between the two mean velocities $v_{dif} = \langle v_{1} \rangle - \langle v_{2} \rangle$ reaches its maximum value.
                The rotated angle of the divider when $v_{dif}$ becomes maximum is the rotation axis of the cluster center projected on the sky, and $v_{rot} = \max \{ v_{dif} \}$ is the rotation amplitude. 
    
                In this study, we apply both methods separately to the Perseus cluster. Also, we divide the data into two groups: member galaxies in $R < 27\arcmin$ and in $27\arcmin < R < 60\arcmin$.
                We choose 27\arcmin\ to make the number of galaxies in each group about the same.
                For this analysis, we assume that the 2D completeness of galaxies does not play a crucial role and we decide to use all galaxies regardless of their brightness. 
                The main justification for this assumption is that the targets observed by MMT/Hectospec were selected only by their apparent magnitudes, and thus our catalog is unbiased when studying global kinematics.
                Figure \ref{fig:rotation} shows the result of applying the Hwang \& Lee method and the Manolopoulou \& Plionis method to the two groups.
                The top and bottom panels are for galaxies in the inner region and outer region, respectively.
                The left panels are the results of the Hwang \& Lee method, with black dots indicating the velocities of individual galaxies and the solid line indicating the fitted sinusoidal curve (Equation \ref{eq:rot_Hwang}).
                The right panels show the results of the Manolopoulou \& Plionis method.
                Although the top left and bottom right panels seem to show a sign of rotation, other results do not.
                The two methods give different results because of their intrinsic characteristic.
                For example, the Hwang \& Lee method assumes that the line-of-sight velocity of a rotating galaxy cluster follows a sinusoidal relation.
                On the other hand, the Manolooulou \& Plionis method may suffer from a binning effect on the position angle, especially when the cluster has a weak rotation signal.
                
                To make a quantitative assessment, we plot the rotational velocity and position angle as a function of distance in Figure \ref{fig:rotation_profile}.
                The top panel shows the rotation amplitude $v_{rot}$ and the bottom panel shows the position angle of the rotation axis, measured from north to east.
                The blue triangles indicate the results following the Hwang \& Lee method, while the red circles represent the results using the Manolopoulou \& Plionis method.
                The error bars are $1\sigma$ resampling uncertainties.
                The ratios between the rotational velocity amplitude and the corresponding uncertainty are 3.12 (inner region) and 1.53 (outer region) for the Hwang \& Lee method and 0.47 (inner region) and 1.86 (outer region) for the Manolopoulou \& Plionis method.
                Although the signal-to-noise ratio for the Hwang \& Lee method in the inner region is relatively high, other results do not present a definite signal of global rotation.

                In addition, we apply a nonparametric test for detecting the rotation. 
                We use a circular-linear correlation test \citep{DirectionalStat}, which tests if there is a correlation between a circular variable and a linear variable.
                A directional variable represents the direction and is in units of degrees or radians.
                That is, 0\arcdeg\ and 360\arcdeg\ are treated as equal values.
                A linear variable does not have this property; examples include velocity and size.
                We utilize the R package \texttt{Directional}, which calculates the $p$ value of the null hypothesis of zero correlation between a directional variable and a linear variable.
                We apply the hypothesis test to the subsamples mentioned above.
                The $p$ values of zero correlation are 0.908 and 0.0001 for the inner region and the outer region, respectively.
                For the inner region, the null hypothesis cannot be rejected even at a 68\% confidence level.
                The null hypothesis for the outer region can be rejected at a 99\% confidence level; however, the existence of a correlation does not necessarily lead to the existence of rotation.
                One counterexample is the case where there is a bulk motion in one part with all other regions showing zero velocity.
                Combining the results from the parametric test and the nonparametric test, it is hard to say for sure that the Perseus cluster is rotating within 60\arcmin.
            
            \subsubsection{Examining the Existence of Substructures}   \label{sec:res:substr}       
\begin{figure*}[htb!]
                    \centering
                    \includegraphics[width=0.6\linewidth]{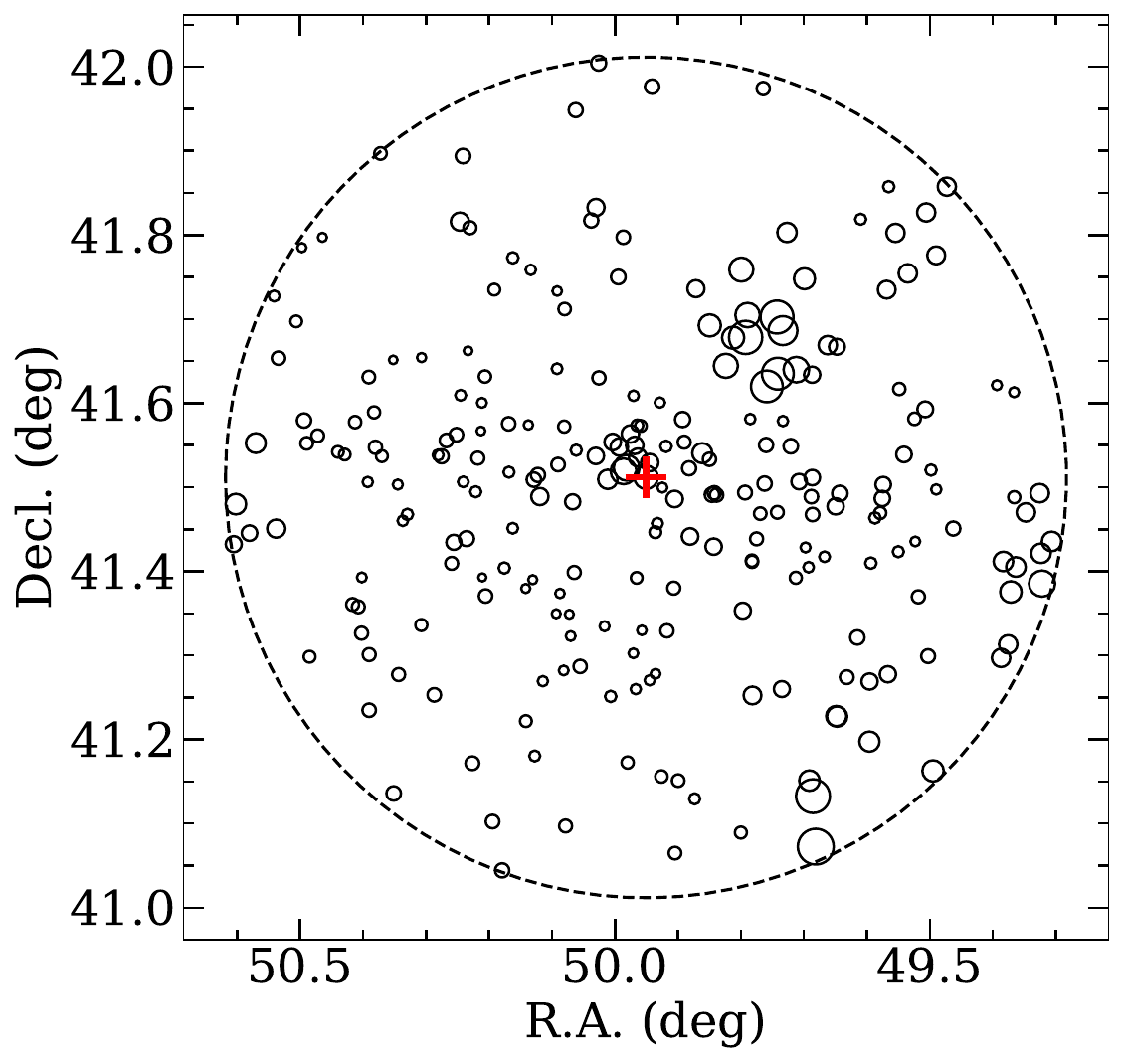}
                    \caption{
                        Distribution of galaxies brighter than 19.1 mag within 30\arcmin.
                        The radius of each circle is proportional to $e^{\delta}$.
                        The cluster center is marked with the red cross.
                        }
                    \label{fig:delta}
                \end{figure*}

                We check for signs of substructure in the Perseus cluster using the $\delta$-test \citep{Dressler...1988AJ.....95..985D}.
                The $\delta$-test uses radial velocities and positional information of galaxies to quantify the local deviation of mean velocity and dispersion from the global values. To be more specific, the $\delta$ value of each galaxy is calculated as 
                \begin{equation}
                    \delta^{2} = (N_{nn}/\sigma^{2}) [ (\bar{v}_{local} - \bar{v})^{2} + (\sigma_{local} - \sigma)^{2}]
                \end{equation}
                where $\bar{v}$ and $\sigma$ denote the mean and the standard deviation of the radial velocities, respectively.
                Variables with the subscript $local$ denote local values and those without subscripts indicate global values.
                The local values are calculated with $N_{nn}$ nearest-neighbors.
                We use $N_{nn}=11$ as suggested by \citet{Dressler...1988AJ.....95..985D}.
                Figure \ref{fig:delta} shows the results of conducting the $\delta$-test with galaxies inside 30\arcmin\ from cluster center and brighter than 19.1 in $r$-band, which are the conditions where the 2D completeness is high and uniform.
                Each circle corresponds to a galaxy, with the radius of the circle proportional to $e^\delta$.
                The outer dotted circle shows the 30\arcmin\ boundary.
                A clump of large circles in the figure would indicate a possible substructure.
                A sparse clump exists at the northwest part of the cluster, which was also noted by \citet{Meusinger...2020AA...640A..30M}.
        
\begin{figure}[htb!]
                    \centering
                    \includegraphics[width=1.0\columnwidth]{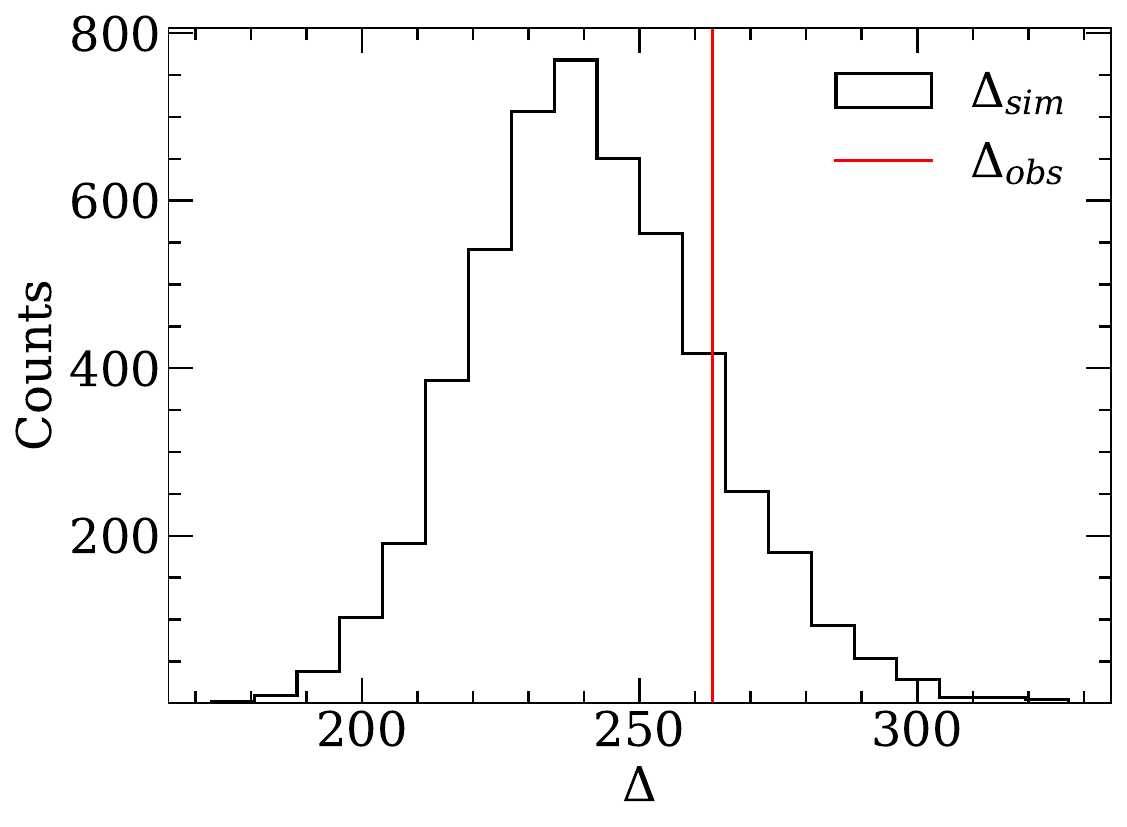}
                    \caption{
                        Histogram of $\Delta$ from Monte-Carlo simulated clusters.
                        The red vertical line indicates the $\Delta$ value of the real cluster.
                        }
                    \label{fig:Delta_hist}
                \end{figure}

                To examine whether such substructures exist, we conduct a Monte-Carlo simulation of the $\Delta$-test \citep{Dressler...1988AJ.....95..985D}, where the $\Delta$ statistic is the sum of all $\delta$.
                We first calculate $\Delta$ of the current data, $\Delta_{obs}$.
                Then, we create 5000 simulated clusters by randomly shuffling the radial velocities of galaxies with the positions fixed and calculate $\Delta$ for each simulated cluster, $\Delta_{sim}$.
                The simulated data will thus have the same positional property and velocity histogram as the observed cluster but with a different combination of the two.
                Figure \ref{fig:Delta_hist} shows the histogram of $\Delta_{sim}$ with the value of $\Delta_{obs}$ marked with the red vertical line.
                The fraction of $\Delta_{sim}$ greater than $\Delta_{obs}$ indicates the significance of the substructure in the observed cluster.
                For our data, $f(\Delta_{sim} > \Delta_{obs})=20\%$, implying that it is statistically insignificant to assert that the current position and velocity distribution are induced from a substructure.
                
    \subsection{Comparison of Spatial Distribution and Kinematics between Gas and Galaxies}    \label{sec:res:icm}
\begin{figure*}[htb!]
            \centering
            \includegraphics[width=1.0\linewidth]{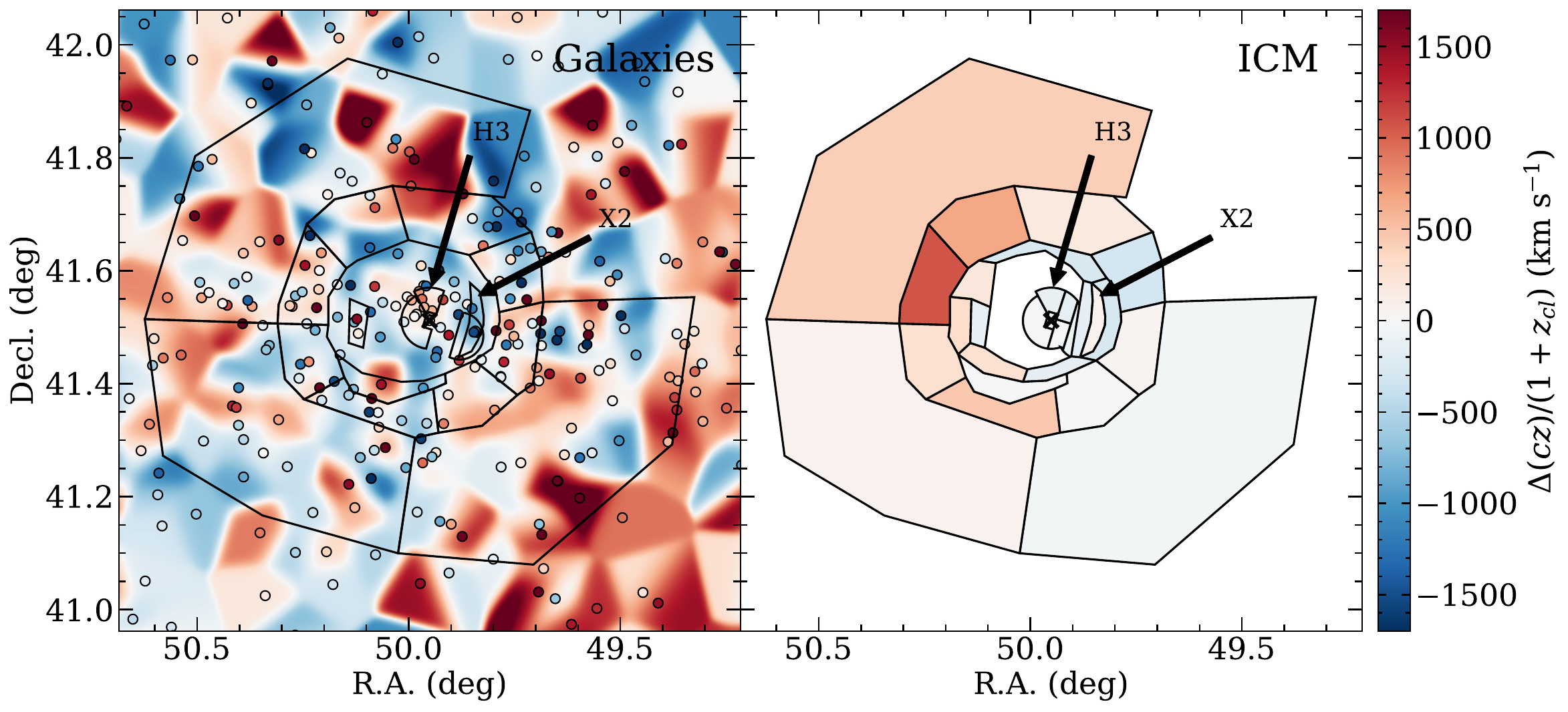}
            \caption{
                Galaxy velocity map (left) and ICM bulk velocity map from \textit{Hitomi} and \textit{XMM-Newton} results (right).
                The galaxy velocity map is weighted with a Gaussian kernel of standard deviation 1.5\arcmin.
                Circles represent individual galaxies color-coded according to their velocities, and the black cross is the center of the cluster.
                }
            \label{fig:hitomi_xmm_map}
        \end{figure*}

        As galaxies and ICM have different timescales for relaxation after a merger (e.g., \citealp{Roetigger...2000ApJ...538...92R}), comparing the kinematics of the two components can reveal some information about the merger history of the cluster. 
        To this end, we compare the velocity maps of galaxies and ICM.
        For the comparison of kinematics, we use all member galaxies regardless of their apparent magnitude.
        We use two data sets for the ICM velocities.
        The first one is the X-ray observation results from the \textit{Hitomi} mission.
        \citet{Hitomi...2018PASJ...70....9H} presented precise ICM velocity measurements in seven central regions of the Perseus cluster.
        Their observation region covers only $R \lesssim 5\arcmin$, so we use another ICM velocity measurement with a wider field from \textit{XMM-Newton} observations.
        Using \textit{XMM-Newton} archive data, \citet{Sanders...2020A&A...633A..42S} measured the velocities of ICM in 21 non-overlapping regions around the center of the Perseus cluster.
        The radius of the total region is $\sim 30\arcmin$, which includes more galaxies compared to the velocity map of \textit{Hitomi}.
        Figure \ref{fig:hitomi_xmm_map} shows the velocity maps of galaxies (left) and ICM (right).
        We mark the region borders with black solid lines.
        For the left panel, we only show the borders within which three or more galaxies exist.
        The \textit{Hitomi} regions lie in the central part, while the \textit{XMM-Newton} regions are located at the outer part.
        The empty region in the \textit{XMM-Newton} results is due to the Cu hole of the \textit{XMM-Newton} detector, which is the region of the detector where energy scale calibration could not be done (refer to \citealp{Sanders...2020A&A...633A..42S} for details).
        We label regions from \textit{Hitomi} (\textit{XMM-Newton}) as `H' (`X') followed by the region number.
        For example, we denote H6 for Region 6 in the \textit{Hitomi} field and X21 for Region 21 in the \textit{XMM-Newton} field.
        We note that H6 and X2 are overlapping with each other, with X2 including only one more member galaxy.
        We compute the galaxy velocity map using member galaxies and a Gaussian weight with a standard deviation of $1.5\arcmin$, which is the narrowest width of the \textit{XMM-Newton} regions.

\begin{figure*}
            \centering
            \includegraphics[width=0.7\linewidth]{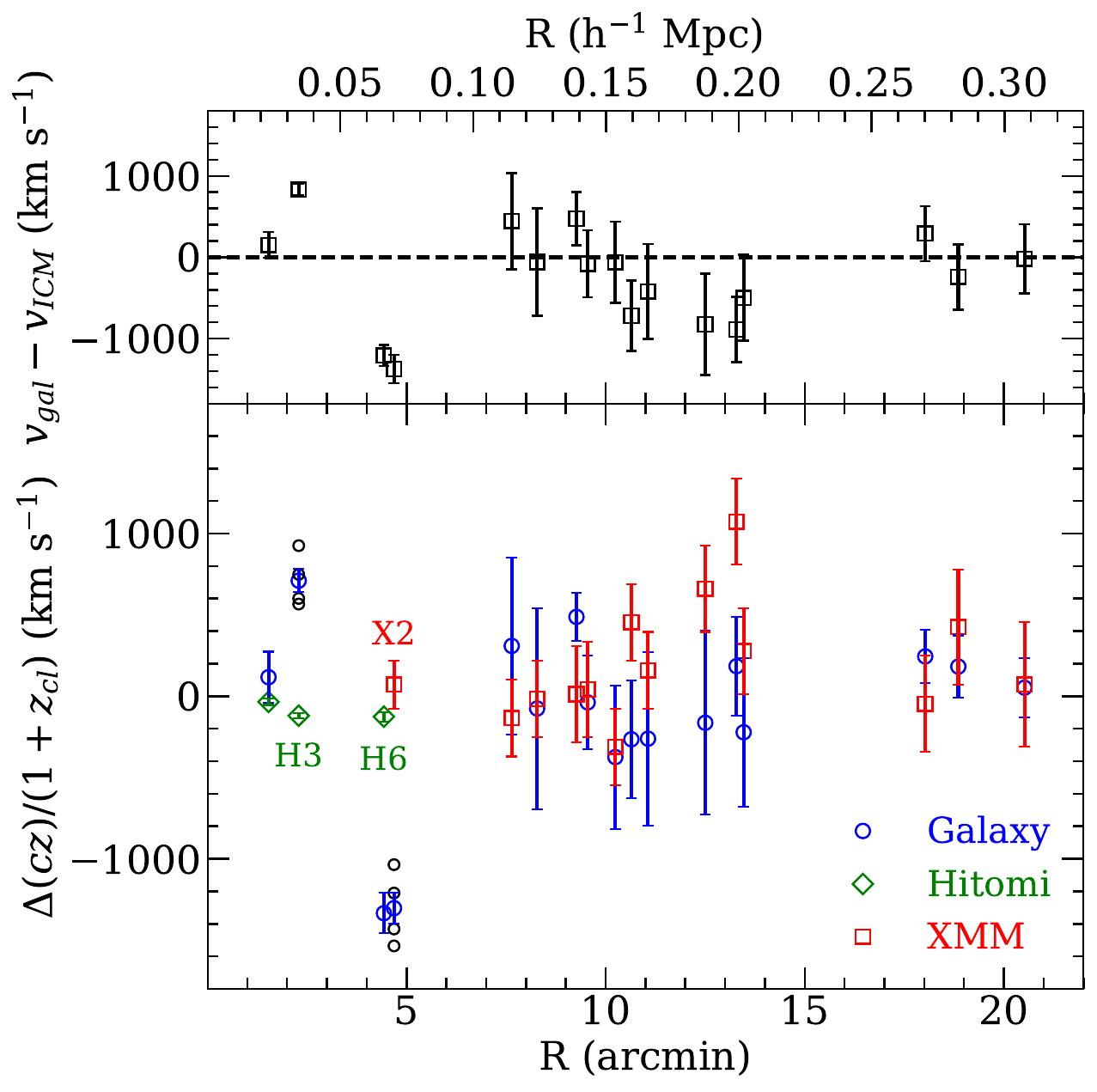}
            \caption{
                Comparison of galaxy velocities and ICM velocities for regions with more than three galaxies.
                (Top) Radial velocity difference of the two components.
                (Bottom) Radial velocities of galaxies (blue open circle), ICM from \textit{Hitomi} results (green filled diamond), and ICM from \textit{XMM-Newton} results.
                The black open circles in H3 and X2 indicate velocities of individual galaxies in the region, demonstrating that a single outlier galaxy in the region does not dominate the difference in the galaxy and ICM velocities.}
            \label{fig:hitomi_xmm_graph}
        \end{figure*}

        For quantitative comparison including error bars, we plot the velocities of the two components as a function of projected distance in Figure \ref{fig:hitomi_xmm_graph}.
        The bottom panel shows the velocities of galaxies (blue open circles), ICM from \textit{Hitomi} (green filled diamonds), and ICM from \textit{XMM-Newton} (red filled squares).
        Only regions with three or more galaxies are plotted.
        The error bars of \textit{Hitomi} data are from Table 4 of \citet{Hitomi...2018PASJ...70....9H} and those of \textit{XMM-Newton} are statistical errors from Table 2 of \citet{Sanders...2020A&A...633A..42S}.
        The velocity errors of galaxies are calculated as $\sigma / \sqrt{N}$, where $\sigma$ is the standard deviation of the velocities of galaxies in the region and $N$ is the number of galaxies in the region.
        The top panel of Figure \ref{fig:hitomi_xmm_graph} shows the velocity difference in each region.
        The error bars are $\sqrt{\sigma_{gal}^{2} + \sigma_{ICM}^{2}}$ where $\sigma_{gal}$ and $\sigma_{ICM}$ are errors of galaxy velocity and ICM velocity, respectively.
        There are three regions with a velocity difference greater than $3\sigma$: H3 ($11.4\sigma$), H6 ($9.5\sigma$), and X2 ($7.8\sigma$).
        As H6 overlaps with X2, there are effectively two regions with significant velocity differences, each located at $R=2.3\arcmin$ and $R=4.7\arcmin$.
        In the bottom plot of Figure \ref{fig:hitomi_xmm_graph}, we plot the radial velocities of individual member galaxies belonging to H3 and X2 (black open circles).
        This confirms that the velocity difference between galaxies and the ICM is not dominated by outlier galaxies in both regions.
        It is also notable that the ICM velocities of the two overlapping regions, H6 and X2, agree well despite the use of different instruments.
        This indicates that the velocity differences between galaxies and ICM in the H3 and X2 are intrinsic features of the cluster, instead of systematic offsets due to instrumental characteristics.
        Other than H3 and X2, the kinematics of galaxies agree with that of the ICM within $3\sigma$.

\begin{figure}
            \centering
            \includegraphics[width=1.0\columnwidth]{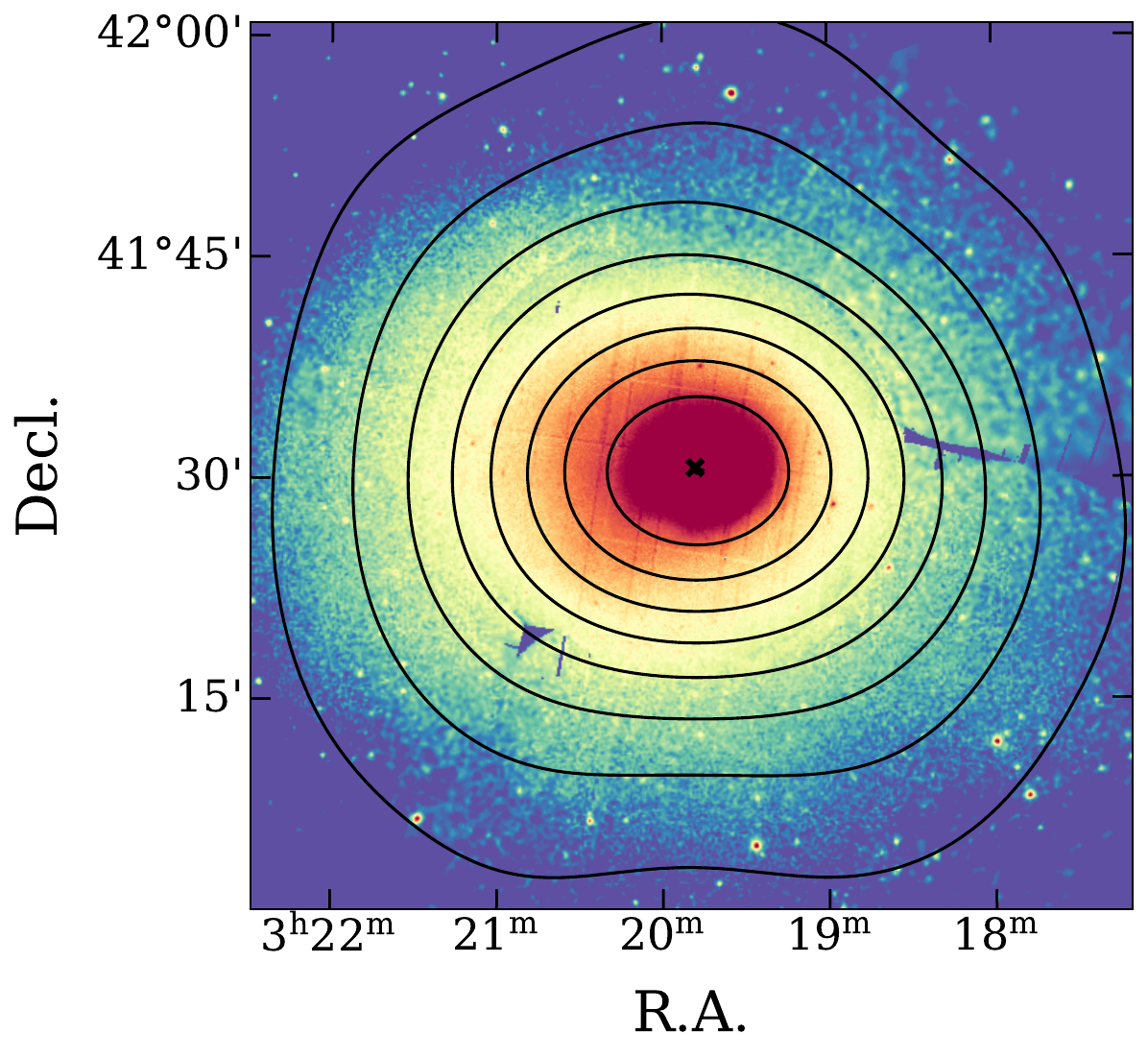}
            \caption{
                Contours of galaxy number density (black line) overlaid on the X-ray intensity map. The galaxy number density is estimated with a Gaussian kernel of standard deviation 10\arcmin.
                }
            \label{fig:numden_overlap}
        \end{figure}

    Lastly, we examine the spatial correlation between galaxy distribution and X-ray emission in $R < 30\arcmin$ region.
    Specifically, we compare the galaxy number density with the X-ray intensity map.
    Galaxy number density is calculated using Gaussian kernel density estimation with the standard deviation of each Gaussian kernel fixed to 10\arcmin.
    We use member galaxies with $r_{\rm{Petro},0} \leq 19.1$.
    For the X-ray intensity map, we use the mosaic image of \textit{XMM-Newton}, retrieved from \textit{XMM-Newton} website\footnote{https://heasarc.gsfc.nasa.gov/docs/xmm/gallery/esas-gallery/xmm\_gal\_science\_perseus.html}.
    Figure \ref{fig:numden_overlap} shows the galaxy number density contours (black line) overlaid on the X-ray intensity map.
    As detector gaps exist in the X-ray image (e.g., empty blanks rightward of the cluster center and bottom-leftward), we resort to qualitative comparison by visual inspection.
    The overall morphology of the cluster traced by ICM and galaxies looks similar; in the inner region ($R \lesssim 15\arcmin$), the X-ray intensity and galaxy number density are elongated in the east-west direction.

\section{Discussion and Conclusions}    \label{sec:discuss_conclusion}
    We compile and present a catalog of galaxies with measured redshifts in the vicinity of the Perseus cluster, consisting of our new data from MMT/Hectospec along with those from the literature.
    We have developed CausticSNUpy, a Python implementation of the caustic technique for cluster membership determination, and analyzed the kinematics of cluster galaxies.
    
    We first note that 1) no confident signal of rotation in $R < 60\arcmin$ is detected, 2) the galaxies in the Perseus cluster do not show substructures within 30\arcmin, and 3) the kinematics and distribution of galaxies and ICM agree at large.
    All three features give hints about past mergers of the cluster.
    For example, the global rotation of a cluster may be a result of a recent off-axis merger \citep{Hwang...2007ApJ...662..236H}.
    Also, substructures within a galaxy cluster are interpreted as infalling galaxy groups \citep{Guennou...2014A&A...561A.112G}.
    The existence of a substructure is evidence of a recent merger and an unrelaxed state.
    Finally, simulations show that galaxies and ICM have different timescales to react to a merger \citep{Roetigger...2000ApJ...538...92R, Shin...2022ApJ...934...43S}.
    In this sense, our results as a whole imply the lack of evidence of a recent merger and suggest that the Perseus cluster is a relaxed cluster.

    There were also previous studies that suggested the relaxation of the Perseus cluster by examining the ICM.
    \citet{Mohr...1993ApJ...413..492M} noted the elongation of ICM isophotes and interpreted it as an X-ray substructure and as a sign of an unrelaxed cluster.
    However, \citet{Markevitch...2001ApJ...562L.153M} reported that relaxed, cool-core clusters can exhibit X-ray substructures.
    Indeed, \citet{Simionescu...2012ApJ...757..182S}, from X-ray image, and \citet{Sanders...2020A&A...633A..42S}, on the basis of low turbulence, saw the Perseus cluster as a relaxed system.
    In this regard, our galaxy-based analysis is consistent with the results from ICM studies.

    It is also interesting that galaxies in the two inner regions ($R<5\arcmin$) have a significant ($> 7\sigma$) difference in velocities from both the ICM and the cluster center (Figure \ref{fig:hitomi_xmm_graph}).
    This difference does not seem to be a global phenomenon at that radius.
    The caustics in Figure \ref{fig:rvmag} tell us that a velocity around 1000km s$^{-1}$ is typical at 5\arcmin.
    \citet{Tamura...2014ApJ...782...38T} noted this deviation of galaxy velocity in X2, along with a similar trend in gas velocities.
    They suggested that this velocity structure is a byproduct of a past minor merger.
    Here, we point out that three among the four galaxies belonging to X2 are very close together when projected on the sky, almost indistinguishable in Figure \ref{fig:hitomi_xmm_map} (dark blue region west of the cluster center).
    The maximum separation between the three galaxies is 20\arcsec, equivalent to a projected distance of 5$h^{-1}$ kpc.
    Thus, the three galaxies in X2 could be a physically bound compact group.
    For H3, the four galaxies are relatively sparsely distributed, with the minimum pair-wise separation being 54\arcsec\ (13$h^{-1}$ kpc).
    The true identities of the velocity offsets in both H3 and X2 need further investigation.
    
    We present a catalog of galaxies with redshift measurements in the central region ($R < 60\arcmin$) of the Perseus cluster.
    The catalog is a product of a deep spectroscopic survey of 1283 galaxies by MMT/Hectospec.
    Among them, we extract 1151 reliable redshifts using the Python code RVSNUpy.
    The catalog consists of 1447 galaxies within 60\arcmin: 1148 from MMT/Hectospec, 265 from SDSS, 29 from NED, and 5 from \citet{Meusinger...2020AA...640A..30M}.
    The new data from MMT/Hectospec greatly increases the spectroscopic completeness in the region.
    Specifically, the \textit{r}-band apparent magnitude limit where the differential completeness drops below 50\% is deepened from 16.1 to 18.0 for $R < 60\arcmin$ and from 16.1 to 19.1 for $R < 30\arcmin$.
    In addition, the spectroscopic completeness is spatially uniform in $R < 30\arcmin$ region for galaxies with $r_{\rm{Petro}, 0} \leq 19.1$.
    We use this new data set to study the kinematics of the Perseus cluster and summarize our primary results as follows.
    \begin{enumerate}
        \item
            We have used CausticSNUpy (an open-source Python module for cluster membership determination) and identified 418 member galaxies within $R < 60\arcmin$. Among them, there are 370 red members and 48 blue members.
        
        \item
            We compute the velocity dispersion profile of member galaxies.
            Future studies could use this data to infer the velocity anisotropy or the mass profile of the cluster.
            Obtaining a stellar velocity dispersion of NGC 1275 would be very helpful for giving a tight constraint on the mass model of the cluster.
            
        \item
            We examine the global rotation of the cluster within 60\arcmin\ from the cluster center but find no promising signal.

        \item
            The Perseus cluster does not exhibit significant substructures confirmed by the $\delta$ and $\Delta$ statistics.

        \item
            The overall kinematics and spatial distributions of galaxies and X-ray-emitting gas agree well on large scales.
            
        \item
            On small scales, the kinematics of galaxies and ICM show significant differences in the inner two regions of the cluster.
            The large velocity difference between the two in region X2 ($\sim 1200 \rm{km\ s^{-1}}$) could be because of a compact galaxy group, while that in region H6 remains uncertain.
            These need to be studied in more detail with more data.
            
    \end{enumerate}
    These signatures imply that the Perseus cluster is a relaxed system and has not experienced a recent merger in the relaxation timescale of galaxies.
    
    Our extensive catalog of galaxies with spectroscopic redshifts will provide a useful data set for studying the dynamics in the central region of the Perseus cluster.
    The utility of our data will be bolstered when coupled with precise X-ray spectroscopic observations in the future, such as XRISM (by Japan Aerospace Exploration Agency; JAXA) and Athena (by European Space Agency; ESA).
    We expect that these high-resolution X-ray missions will widen the observed region by \textit{Hitomi}, which was only operable for about a month \citep{Hitomi...2016Natur.535..117H}.

% \begin{acknowledgments}
\section*{Acknowledgments}
    This work was supported by K-GMT Science Program (PID: UAO-G181-21B) funded through Korean GMT Project operated by Korea Astronomy and Space Science Institute (KASI). Observations reported here were obtained at the MMT Observatory, a joint facility of the University of Arizona and the Smithsonian Institution. HSH acknowledges the support by the National Research Foundation of Korea (NRF) grant funded by the Korea government (MSIT) (No. 2021R1A2C1094577). 
    HS acknowledges the support of the National Research Foundation of Korea (NRF) grant funded by the Korean government (MSIT) (No. 2022R1A4A3031306).
    NH and BGP acknowledge the support by the Korea Astronomy and Space Science Institute (KASI) grant funded by the Korean government (MSIT; No. 2023-1-860-02, International Optical Observatory Project). Funding for the Sloan Digital Sky Survey IV has been provided by the Alfred P. Sloan Foundation, the U.S. Department of Energy Office of Science and the Participating Institutions. SDSS-IV acknowledges support and resources from the Center for High-Performance Computing at the University of Utah. The SDSS web site is [www.sdss.org](http://www.sdss.org/).
    
    SDSS-IV is managed by the Astrophysical Research Consortium for the Participating Institutions of the SDSS Collaboration including the Brazilian Participation Group, the Carnegie Institution for Science, Carnegie Mellon University, the Chilean Participation Group, the French Participation Group, Harvard-Smithsonian Center for Astrophysics, Instituto de Astrofísica de Canarias, The Johns Hopkins University, Kavli Institute for the Physics and Mathematics of the Universe (IPMU) / University of Tokyo, Lawrence Berkeley National Laboratory, Leibniz Institut für Astrophysik Potsdam (AIP), MaxPlanck-Institut für Astronomie (MPIA Heidelberg), Max-PlanckInstitut für Astrophysik (MPA Garching), Max-Planck-Institut für Extraterrestrische Physik (MPE), National Astronomical Observatories of China, New Mexico State University, New York University, University of Notre Dame, Observatário Nacional / MCTI, The Ohio State University, Pennsylvania State University, Shanghai Astronomical Observatory, United Kingdom Participation Group, Universidad Nacional Autónoma de México, University of Arizona, University of Colorado Boulder, University of Oxford, University of Portsmouth, University of Utah, University of Virginia, University of Washington, University of Wisconsin, Vanderbilt University and Yale University.
% \end{acknowledgments}

\appendix
\section{Caustic Technique} \label{sec:app:caustic}
    \subsection{Process of the Caustic Technique}
    Here, we give an overview of the caustic technique described by \citet{Diaferio...1999MNRAS.309..610D} and \citet{Serra...2011MNRAS.412..800S}.
    First, candidate members of the cluster are selected using a hierarchical clustering algorithm.
    The algorithm iteratively clusters galaxies into groups based on their 2D positions and velocities.
    More specifically, the similarity between two galaxies $i, j$ are defined by their binding energy
    \begin{equation}
        E_{ij} = -G\frac{m_{i}m_{j}}{R_{p}} + \frac{1}{2} \frac{m_{i} m_{j}}{m_{i} + m_{j}} \Pi^{2}
    \end{equation}
    where $m_{i}$ and $m_{j}$ are the mass of  galaxy $i, j$ (fixed to $10^{12} h^{-1} \mathrm{M_{\Sun}}$), $R_{p}$ is the projected separation, and $\Pi$ is the line-of-sight velocity difference.
    At first, each individual galaxy is a group by itself.
    At each step, the hierarchical clustering algorithm merges two groups with the highest similarity (i.e., the strongest binding energy).
    The similarity between group $g_{\alpha}$ and $g_{\beta}$ are defined as
    \begin{equation}
        E_{\alpha \beta} = \min_{i \in g_{\alpha}, j \in g_{\beta}} E_{i,j}
    \end{equation}
    which is the minimum pairwise binding energy between $g_{\alpha}$ and $g_{\beta}$.
    The hierarchical clustering iterates this grouping until all galaxies are in a single group.
    
    A dendrogram, also called a binary tree, showing the grouping history can be constructed after the hierarchical clustering is done.
    The path from the root to the leaves that follows the subtree with the most leaves hanging from each junction is called the main branch.
    To shortlist the candidate members from the binary tree, we need to cut the main branch at a certain node and choose the galaxies hanging from that node.
    The criterion for selecting the threshold node is the velocity dispersion of galaxies hanging from the node.
    During hierarchical clustering, when a substructure or an interloper is grouped into the main branch, the velocity dispersion would increase significantly.
    Otherwise, the velocity dispersion maintains a constant value.
    Using this property, the threshold node is chosen as the point on the main branch where the velocity dispersion of galaxies changes abruptly.
    The detailed implementation of detecting this change differs from \citet{Diaferio...1999MNRAS.309..610D} and \citet{Serra...2011MNRAS.412..800S}.
    Note that these candidate members do not necessarily coincide with the final members determined using the caustics.
    
    If the cluster center is not given, it can be estimated using the candidate members.
    The right ascension and declination of the cluster center are determined as the peak of the number density of candidate members on the 2D sky $(\alpha, \delta)$.
    The number density is calculated using kernel density estimation.
    The median of the candidate members is used as the redshift of the cluster center.
    
    Next, we draw the redshift diagram using all galaxies.
    The Hubble parameter $H_0$ is used to convert the projected distance to have the same physical dimension as the velocity.
    Another parameter, $q$, is used to rescale the axes and adjust the weights of each axis.
    \citet{Diaferio...1999MNRAS.309..610D} chooses to use $q=25$ to match the typical uncertainties in the measured positions and measured velocities.
    Using these two parameters, we can plot the candidate member galaxies in the redshift space as $\mathbf{x}=(rH_{0}q, v)$ where $r$ is the projected distance of a galaxy from the cluster center and $v$ is the radial velocity relative to the cluster center.

    We then estimate the number density of galaxies in the redshift space.
    The galaxy number density $f_{q}(\mathbf{x})$ is estimated using adaptive kernel density estimation, which gives different kernel sizes to galaxies.
    The number density at point $\mathbf{x}$ is estimated as
    \begin{equation}
        f_{q}(\mathbf{x}) = \frac{1}{N} \sum_{i=1}^{N} \frac{1}{h_{i}^{2}} K(\frac{\mathbf{x}-\mathbf{x}_{i}}{h_{i}})
    \end{equation}
    where $N$ is the number of galaxies, $\mathbf{x}_{i}$ is the redshift space coordinates of each galaxies, $K$ is the kernel, and $h_{i}$ is the adaptive kernel width assigned to each galaxies.
    \citet{Diaferio...1999MNRAS.309..610D} chose to use the triweight function as a kernel defined as
    \begin{equation}
        K(\mathbf{x}) = \begin{cases}
                        \frac{4}{\pi} (1-| \mathbf{x} |^{2})^{3} ,   & |\mathbf{x}| < 1 \\
                        0                                        ,   & \text{otherwise}
                        \end{cases}
    \end{equation}
    The determination of adaptive kernel width is described in detail by \citet{Diaferio...1999MNRAS.309..610D}.
    
    After estimating the galaxy number density, we determine the caustic amplitude.
    Ideally, the caustics would be the solution $f_{q}(r, v)=0$.
    In real situations, however, there exist interlopers and escaping galaxies.
    Thus, a threshold $\kappa$ is used so that the contour $f_{q}(r, v) = \kappa$ determines the caustic location.
    In addition, the contour $f_{q}(r, v) = \kappa$ would not generally be symmetric along $v=0$ in the redshift space.
    Because spherical symmetry is assumed, we make the caustics symmetric.
    We decompose the solution of $f_{q}(r, v) = \kappa$ into the upper part $v_{up}(r)$ ($v>0$) and the lower part $v_{low}(r)$ ($v<0$).
    The upper caustics $\mathcal{V^{+}}(r)$, the lower caustics $\mathcal{V^{-}}(r)$, and in turn the caustic amplitude $\mathcal{A}(r)$ are then chosen as
    \begin{equation}    \label{eq:caustics}
        \mathcal{V}^{\pm}(r) = \pm\min\{|v_{up}(r)|, |v_{low}(r)|\}
    \end{equation}
    \begin{equation}    \label{eq:caustic_amp}
        \mathcal{A}(r) = \frac{1}{2} [\mathcal{V}^{+}(r) - \mathcal{V}^{-}(r)].
    \end{equation}
    In addition, simulated galaxy clusters show $d \ln \mathcal{A} / dr \lesssim 1/4$.
    To reflect this fact, a point on the caustic lines with a logarithmic derivative greater than $\zeta $ is replaced with a value that satisfies  $d \ln \mathcal{A} / dr = 1/4$.
    \citet{Diaferio...1999MNRAS.309..610D} chose $\zeta=1$ while \citet{Serra...2011MNRAS.412..800S} used $\zeta=2$.
    The final cluster members are those within the caustics.
       
    One task remains, which is choosing the right threshold $\kappa$.
    The caustics correspond to escape velocity at the given distance, thus $\kappa$ can be chosen to minimize 
    \begin{equation} \label{eq:optimize}
        S(\kappa, R) = | \langle v^{2}_{esc} \rangle_{\kappa, R} - 4 \langle v^{2} \rangle_{R} | ^{2}
    \end{equation}
    where $R$ is the mean projected distance of candidate members, $ \langle v^{2}_{esc} \rangle_{\kappa, R} = \int^{R}_{0} \mathcal{A}^2(r) \phi(r)dr / \int^{R}_{0} \phi(r) dr$ is the mean squared caustic amplitude within $R$, and $\phi(r) = \int f_{q}(r, v) dv$.
    Although $S(\kappa)=0$ indicates that the cluster is in dynamical equilibrium within the radius $R$, \citet{Serra...2011MNRAS.412..800S} assert that this optimization gives valid $\kappa$ regardless of the dynamical state of the cluster.

    The caustic technique can be summarized as the following steps:
    \begin{enumerate}
        \item Find candidate cluster members using a hierarchical clustering algorithm.
        \item If necessary, find the cluster center by locating the peak of the galaxy number density and calculating the mean of the candidate member redshifts.
        \item Draw the redshift diagram using all galaxies and rescale the two axes using $H_0$ and $q$ so that they have the same units and are evenly weighted.
        \item Estimate the galaxy number density $f_{q} (\mathbf{x})$ using adaptive kernel density estimation.
        \item Start with an arbitrary $\kappa$.
        \item For given $\kappa$, find the solution to $f_{q}(r, v) = \kappa$.      \label{step:contour}
        \item Calculate the caustic amplitude from Equations (\ref{eq:caustics}) and (\ref{eq:caustic_amp}).
        \item Evaluate Equation (\ref{eq:optimize}).        \label{step:Sk}
        \item Change the value of $\kappa$ and repeat steps \ref{step:contour} through \ref{step:Sk} until a minimum has been found.
        \item Final member galaxies are selected as those within the caustics.
    \end{enumerate}

    \subsection{CausticSNUpy and Caustic App}

    \begin{figure*}[htb!]
        \centering
        \includegraphics[width=1.0\linewidth]{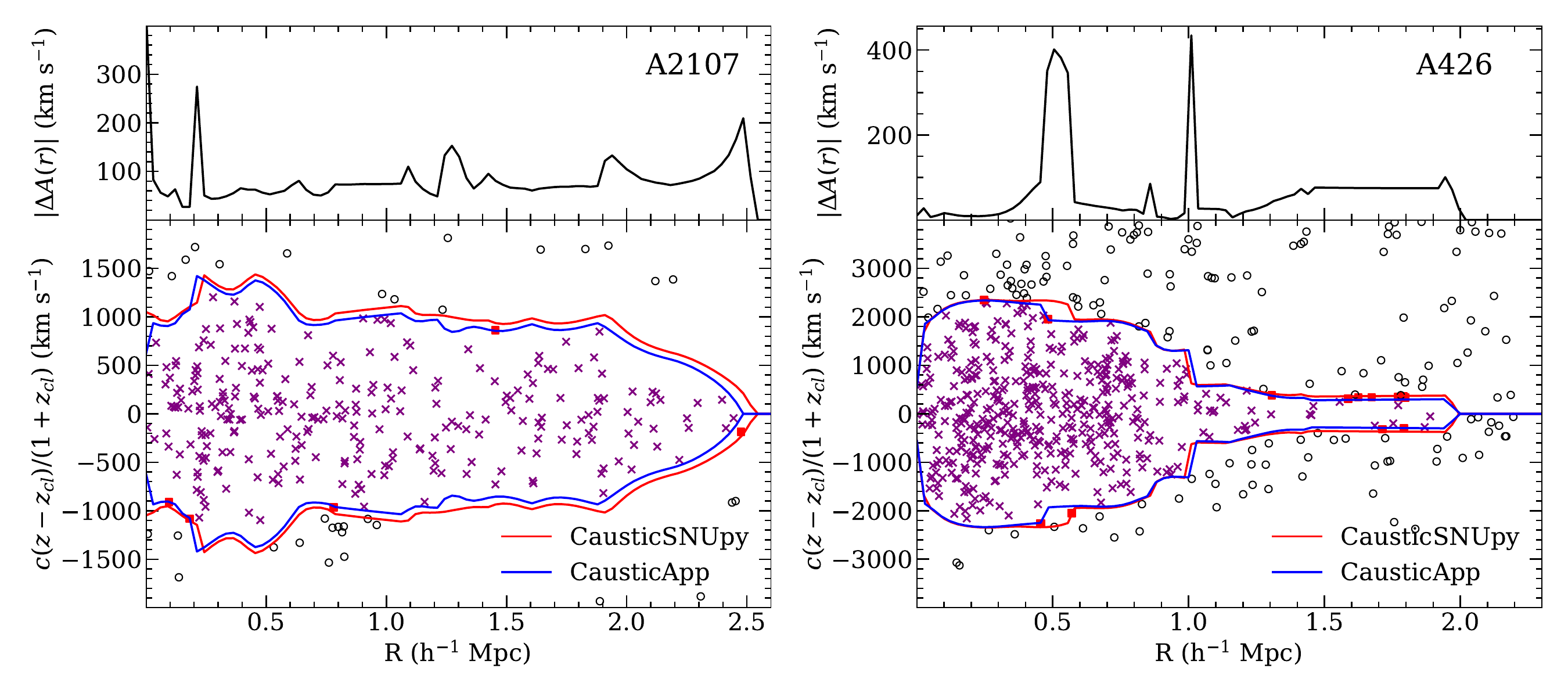}
        \caption{
            Comparison of the results by our CausticSNUpy and the traditional Caustic App.
            (Top) Absolute difference in the caustic amplitude.
            (Bottom) Redshift diagram showing the caustics determined by CausticSNUpy (red line) and Caustic App (blue line).
            Purple crosses are galaxies identified as cluster members by both CausticSNUpy and Caustic App.
            Red squares and blue triangles are galaxies identified only by CausticSNUpy and Caustic App, respectively.
            Black open circles are non-members as identified by both programs.
        }
        \label{fig:caustic_compare}
    \end{figure*}
    
    We test the consistency of the caustics determined by our code, CausticSNUpy, to those from the existing software Caustic App.
    As our aim is to implement the caustic technique described by \citet{Diaferio...1999MNRAS.309..610D} and \citet{Serra...2011MNRAS.412..800S}, we hereinafter regard the result of Caustic App as the answer.
    We use two data sets for comparison: the catalog presented by \citet{Song...2017ApJ...842...88S} for galaxy cluster Abell 2107 (A2107) and the catalog in this paper for the Perseus cluster (A426).
    The bottom panels of Figure \ref{fig:caustic_compare} show the result of the caustic technique by CausticSNUpy (red) and Caustic App (blue).
    The purple crosses represent galaxies identified as members by both CausticSNUpy and Caustic App.
    The red squares and the blue triangles are the galaxies determined as members by CausticSNUpy but not by Caustic App and vice-versa.
    The top panels show the difference in the caustic amplitudes calculated by the two programs.
    Qualitatively, the caustics calculated by our CausticSNUpy have an overall similar shape as those yielded by the existing Caustic App.
    
    For quantitative assessment, we calculate the completeness and excess of member determination compared to the Caustic App.
    For A2107, CausticSNUpy identifies all member galaxies found by Caustic App but also includes 5 more galaxies; the completeness is 100\% with 1.8\% excess identification.
    For the Perseus cluster, the member galaxies identified by CausticSNUpy include 12 more galaxies while finding all member galaxies identified by Caustic App.
    Thus, the completeness of CausticSNUpy is 100\% while the excess identification rate is 6.8\% for the Perseus cluster.
    We also observe for both programs that the caustic location changes noticeably when different settings are used, such as the initial velocity cuts applied before starting the hierarchical clustering.
    In addition, as \citet{Diaferio...1999MNRAS.309..610D} noted, the choice of $\kappa$ as the threshold of number density can be adjusted manually.
    Taking these factors into account, CausticSNUpy can act as a reliable open-source Python code to run the caustic technique and as a substitute for Caustic App.

\bibliography{A426}{}
\bibliographystyle{aasjournal}

\end{document}